%% file: main.tex
\documentclass[conference]{IEEEtran}

\usepackage{cite}
\usepackage{amsmath,amssymb,amsfonts}
\usepackage{algorithmic}
\usepackage{graphicx}
\usepackage{textcomp}
\usepackage{xcolor}
\usepackage{url}
\usepackage{enumitem}
\usepackage{tabularx,booktabs,ragged2e}
\usepackage{tikz}
\usetikzlibrary{shapes,arrows.meta,positioning}
\usepackage{float}
\usepackage[T1]{fontenc}
\usepackage{multirow}
\usepackage{makecell}
\usepackage{subcaption}
\usepackage{hyperref}

\newcommand{\good}[1]{\textcolor{green!60!black}{\scalebox{1.2}{$\downarrow$}}\,#1}
\newcommand{\bad}[1]{\textcolor{red}{\scalebox{1.2}{$\uparrow$}}\,#1}
\usepackage{hyperref}

\ifCLASSINFOpdf
\else
\fi

\begin{document}
\pagestyle{plain}

\title{Breaking Windows Malware Detection: A Comprehensive Evaluation of Problem-Space Adversarial Robustness}

\author{
\begin{tabular}{ccc}
\begin{minipage}{0.30\textwidth}
\centering
\textbf{Mashal Zainab}\\
\textit{University of Luxembourg}\\
mashal.zainab@uni.lu
\end{minipage}
&
\begin{minipage}{0.30\textwidth}
\centering
\textbf{Salijona Dyrmishi}\\
\textit{Independent Researcher}\\
dyrmishisalijona@gmail.com
\end{minipage}
&
\begin{minipage}{0.30\textwidth}
\centering
\textbf{Hamid Bostani}\\
\textit{University of Luxembourg}\\
hamid.bostani@uni.lu
\end{minipage}
\\[1.3em]
\multicolumn{3}{c}{
\begin{tabular}{cc}
\begin{minipage}{0.36\textwidth}
\centering
\textbf{Lorenzo Cavallaro}\\
\textit{University College London}\\
l.cavallaro@ucl.ac.uk
\end{minipage}
&
\begin{minipage}{0.36\textwidth}
\centering
\textbf{Maxime Cordy}\\
\textit{University of Luxembourg}\\
maxime.cordy@uni.lu
\end{minipage}
\end{tabular}
}
\end{tabular}
}


\maketitle

\begin{abstract}

Problem-space evasion attacks have exposed critical weaknesses in machine learning-based malware detectors; yet, their evaluation remains fragmented across models, datasets, and attack methodologies, often neglecting domain-specific requirements such as executability and functionality preservation. We address this gap with a unified, large-scale evaluation of nine state-of-the-art evasion attacks against eight Windows malware detectors, including seven open-source models and one commercial detector, under executability-preserving conditions. Our study analyzes attack effectiveness, complementarity, transferability, and adversarial hardening to evaluate robustness along complementary dimensions. We show that detector vulnerability depends strongly on both model representation and attack type: raw-byte detectors are particularly susceptible to several classes of problem-space manipulation, but no detector family is uniformly robust across all attacks. Importantly, effectiveness is not explained by transformation-space size alone: the strongest attacks can achieve substantially higher success while using fewer distinct transformations and concentrating on a small set of high-impact manipulations. We further show that two complementary attacks are sufficient to cover $\approx 99\%$ of the adversarial examples produced by the remaining evaluated attacks. Transferability exhibits a different pattern from direct attack success: attacks with low direct success can produce highly transferable evasions. Finally, adversarial hardening is highly attack- and model-dependent: robustness gains often fail to transfer across attacks and can even increase susceptibility to unseen attacks. These findings highlight important limitations in current malware robustness evaluation practices, establish a comprehensive empirical baseline for realistic evasion attacks, and clarify key relationships between effectiveness, transferability, and defense robustness.

\end{abstract}


\IEEEpeerreviewmaketitle

\input{01_introduction}
\input{02_background}

\input{03_related_work}

\input{05_rqs}

\input{06_exp_setting}

\section{Results}
\label{sec:results}
\input{rq1}
\input{rq2}

\input{rq3}

\input{rq4}

\input{08_discussion}
\input{09_conclusion}

\bibliographystyle{IEEEtran}
\bibliography{references}

\input{Appendix}

\end{document}

%% file: 01_introduction.tex
\section{Introduction}
\label{sec:intro}

As Windows malware continues to pose a major threat to modern computing systems, machine learning (ML) has emerged as a promising defense mechanism, providing the ability to generalize beyond traditional signature-based detection, and analyze large volumes of files at scale. Despite their widespread adoption, these systems operate in a fundamentally adversarial setting. Malware authors actively develop problem-space evasion attacks, methods that generate adversarial malware variants by modifying real portable executables (PEs) files in a way that preserves functionality while making these variants incorrectly classified as benign. As a result, even well-performing ML-based detectors can be vulnerable to these carefully crafted samples, raising serious concerns about their reliability and robustness in real-world scenarios.

In response to the growing sophistication of adversarial malware, the security community has pursued two primary directions: devising new evasion attacks to probe classifier weaknesses~\cite{demetrio2021functionality,song2020automatic,sharif2019optimization,khormali2019copycat,rosenberg2020query, li2025minimal, zhan2023malpatch, liugrasp},  and constructing new defenses to fortify them~\cite{dyrmishi2023empirical,lucas2023adversarial,bostani2022domain,doan2023feature,bostani2024effectiveness,li2023pad}. Yet, the practical efficacy of these efforts remains unclear, as they are evaluated within a fragmented and inconsistent landscape \cite{bensaoud2024survey,aryal2024survey}. This lack of a standardized benchmark means that new evasion attacks are often validated against only a handful of detectors (e.g.,~\cite{song2020mab}), while new defenses are tested against a narrow and often impractical set of evasion attacks (e.g.,~\cite{li2023pad}). Consequently, this fragmentation hinders progress by making it impossible to identify the most potent evasion attacks or the most effective defenses, ultimately leaving systems vulnerable. This state of uncertainty forces us to confront a critical question:



\noindent\textit{How accurately do current evaluations capture the adversarial robustness of malware classifiers, and how reliable are the conclusions drawn from them?}

Answering this question is nontrivial. Evaluating adversarial robustness in Windows malware detection requires more than measuring whether an attack changes a model prediction. Problem-space attacks must produce valid executable binaries. At the same time, large-scale evaluation remains difficult due to the limited availability of reproducible datasets and models, as well as the computational cost of problem-space attacks. These challenges have constrained the scope of prior evaluations and can lead to robustness conclusions that do not generalize across attacks or detectors, often resulting in narrow or over-optimistic robustness claims.

To address these challenges, we conduct a large-scale empirical study of adversarial robustness in Windows malware detection under a common evaluation methodology. We develop an evaluation framework that jointly considers dataset construction, detector representation, attack strategy, runtime validity, and
cross-model evaluation. Our experimental testbed comprises eight malware classifiers and nine problem-space evasion attacks, spanning raw-byte, image-based, and feature-engineered detectors, as well as attack strategies with substantially different optimization and transformation mechanisms. Rather than evaluating attacks in isolation, we study how their effectiveness, overlap, transferability, and induced robustness interact across detectors.

Building on this framework, we investigate four research questions
(Sec.~\ref{sec:rqs}) concerning (i) the effectiveness of black-box problem-space evasion attacks, (ii) the complementarity and coverage of different attack strategies, (iii) cross-model transferability, and (iv) the effectiveness and cross-attack generalization of adversarial hardening. Our results reveal four main findings. First, attack effectiveness is strongly dependent on both the target detector and attack strategy: no single attack uniformly characterizes the vulnerability of all evaluated detectors, and transformation-space size alone does not explain attack strength. Second, complementary attacks expose adversarial examples missed by the strongest individual attack, and a compact two-attack ensemble covers approximately 99\% of the evasive examples observed within our benchmark. Third, direct attack effectiveness and transferability are distinct properties: attacks with limited direct success can still produce highly transferable evasions, while transfer success depends strongly on the surrogate--target pairing. Finally, adversarial hardening is highly model- and attack-dependent. Robustness learned against one attack often generalizes unevenly to others and can, in some cases, increase susceptibility to previously unseen attacks and does not necessarily provide comparable protection against other, even closely related, attacks. 

In summary, we make the following contributions:

\begin{enumerate}

    \item We develop a unified evaluation methodology for studying adversarial robustness in Windows malware detection across direct attack, attack complementarity, cross-model transfer, and adversarial hardening settings. Our implementation is freely available and can be used by future research to increase the rigor of robustness evaluation. \footnote{https://github.com/mashalzainab/windows-malware-adversarial-robustness/}    
    \item We conduct a large-scale empirical evaluation of eight malware classifiers and nine problem-space evasion attacks, covering raw-byte, image-based, and feature-engineered detectors together with diverse attack strategies. To the best of our knowledge, this is the first study to jointly examine such a broad range of evasion attacks and detection paradigms. In contrast to prior evaluations centered primarily on individual attacks or robustness dimensions, our study jointly examines attack effectiveness, overlap, transferability, and cross-attack defense
    generalization. 
    \item We identify four key empirical insights concerning attack effectiveness, transferability, attack complementarity, and adversarial hardening. These findings provide concrete recommendations for designing more reliable robustness evaluations and for interpreting adversarial vulnerability and robustness claims in Windows malware detection. 

\end{enumerate}

%% file: 02_background.tex
\section{Background}
\label{background}

\subsubsection{Evasion Attacks} 
Malware classifiers are vulnerable to adversarial attacks \cite{biggio2013evasion}, in which adversaries craft subtle modifications to malicious code to evade ML-based detectors at inference-time \cite{goodfellow2014explaining,kolosnjaji2018adversarial,aryal2024survey}, which are trained to distinguish between benign and malicious samples. These manipulations, known as evasion attacks, deliberately obfuscate malware to bypass detection while preserving its malicious functionality and undermine the trustworthiness of malware defense systems.

\noindent\textbf{Problem-Space vs. Feature-Space Attacks.} 
Evasion attacks are commonly categorized into feature-space and problem-space attacks. Feature-space attacks model inputs in abstract feature vectors and manipulate the abstract feature representations used by the classifier, often without ensuring that the resulting samples remain realistic and functional. In contrast, problem-space attacks operate on executable malware, applying realizable transformations that preserve functionality, semantics, and plausibility \cite{pierazzi2020intriguing}, and therefore provide a realistic threat model for deployed systems \cite{demetrio2021functionality}. Because problem-space attacks correspond to actual manipulations that an adversary could perform, they provide a more realistic and security-relevant evaluation of the robustness of Windows malware detectors.

\noindent\textbf{White-Box vs. Black-Box Attacks.} 
Evasion attacks can be further divided into white-box \cite{kolosnjaji2018adversarial,demetrio2019explaining,demetrio2021adversarial} and black-box \cite{anderson2018learning,song2020mab,labaca-castro2021aimed-rl,yuan2020black,zhong2023malfox} depending on their access to or knowledge of the internal structure or parameters of the target model. White-box attacks assume full model knowledge (gradients, weights) and enable gradient-guided methods, while black-box attacks operate with limited feedback (labels or scores) and better reflect real-world attacker capabilities.

Accordingly, this work focuses on black-box problem-space attacks, which provide a realistic assessment of malware detectors under the limited knowledge and operational constraints typical of deployed systems.

\subsubsection{Static Windows Malware Classifiers}
Malware classifiers have been extensively studied using either static or dynamic features. Static analysis extracts information from an executable without running it, for example, by examining headers, strings, byte sequences, or control-flow structures. Dynamic analysis, in contrast, observes program behavior during execution, such as API calls, system interactions, or network activity, typically within a controlled sandbox environment. Static techniques are generally more efficient, provide broader code coverage \cite{chen2021malware}, and can scale to millions of files without executing potentially harmful code, making them a practical first line of defense in real-world deployments.
Static Windows malware classifiers can be broadly categorized into four groups based on their feature representation \cite{ucci2019survey,gibert2024machine}: engineered (tabular) features, visualization-based methods, end-to-end raw-byte models, and graph-based approaches.

\textbf{Engineered feature-based} techniques rely on statically extracted tabular features, such as section entropy, imported functions, byte histograms, and metadata. A notable contribution in this category is the EMBER dataset \cite{anderson2018ember,joyce2025ember2024}, which has become a widely used benchmark and feature extraction standard. These approaches are efficient, interpretable, and remain relevant, though they depend on the quality of hand-crafted features.

\textbf{Visualization-based} techniques convert binaries into images for classification. \cite{nataraj2011malware} proposed one of the first grayscale image representations of executables, a method that has inspired numerous follow-up studies \cite{kancherla2013image,go2020visualization,vinayakumar2019robust,seneviratne2022self,prima2020using,bhodia2019transfer}. By exposing spatial patterns in the byte sequences, these methods enable convolutional neural networks to identify malware-specific textures effectively.

\textbf{End-to-end learning-based} techniques operate directly on raw executable bytes, avoiding manual feature engineering. \cite{raff2017malware} introduced MalConv, a CNN trained on raw PE files, and \cite{raff2021classifying} later improved it to better handle the extreme length of executables. These methods leverage the representational power of deep neural networks, but must manage computational challenges due to sequence length.

\textbf{Graph-based} techniques represent executables as structured graphs, such as Function Call Graphs (FCGs), Control Flow Graphs (CFGs), or Program Dependence Graphs (PDGs), and apply graph representation learning (GRL). For example, MAGIC \cite{yan2019classifying} converts assembly code into CFGs and applies a Deep Graph Convolutional Neural Network (DGCNN) \cite{zhang2018end}, while MalGraph \cite{ling2022malgraph} uses hierarchical FCGs and CFGs to capture structural semantics. By modeling relationships between functions and instructions, graph-based methods can identify malware patterns that may be invisible to simpler representations.

All these approaches face robustness challenges. Engineered features, while efficient and interpretable, can be bypassed by attacks targeting specific features. Visualization and raw-byte methods reduce manual feature design but remain vulnerable to subtle binary perturbations. Graph-based techniques capture richer program semantics but can still be manipulated through structural changes. 

In this work, we focus on static malware detectors and systematically evaluate a diverse set of detectors spanning the major categories to assess their robustness against problem-space evasion attacks.

\subsubsection{Black-Box Problem-Space Windows Malware Evasion Attacks}
In the Windows ecosystem, problem-space evasion attacks operating under black-box settings can be categorized by their search strategies \cite{ling2023adversarial}:


\textbf{Reinforcement learning-based} attacks iteratively select sequences of semantic-preserving modifications to evade detection. Examples include Gym-malware \cite{anderson2017evading,anderson2018learning}, MAB-malware \cite{song2020mab}, AIMED-RL \cite{labaca-castro2021aimed-rl}, MalwareTotal \cite{he2024malwaretotal}, AMG-VAC \cite{ebrahimi2021binary}, DQEAF \cite{fang2019evading}, SRL \cite{zhang2022semantics}, MalInfo \cite{zhong2022reinforcement}, gym-malware-mini \cite{chen2020generating}, and \cite{gibert2022enhancing} which employ RL agents or actor-critic variants to learn effective modification policies.

\textbf{Randomization-based} attacks apply stochastic transformations to the malware binary, such as section injection, code padding, or embedding into a dropper. ARMED \cite{labaca-castro2019armed} and Dropper \cite{ceschin2019shallow} are representative methods, using random modifications while preserving functionality.

\textbf{Evolutionary or Population or Genetic algorithm-based} attacks generate a population of candidate adversarial binaries and evolve them over generations using genetic operations such as selection, crossover, and mutation. GAMMA \cite{demetrio2021functionality}, AIMED \cite{labaca-castro2019aimed}, MDEA \cite{wang2020mdea}, and AMG-PDG \cite{wang2022black} fall into this category.

\textbf{Generative adversarial network-based} attacks leverage generative adversarial networks (GANs) to produce adversarial payloads or perturbations that can be appended to the malware. GAPGAN \cite{yuan2020black}, MalFox \cite{zhong2023malfox} and \cite{rosenberg2020query} exemplify this approach.

\textbf{Heuristic or search-based} attacks employ targeted modifications guided by hill climbing, occlusion analysis, or other deterministic strategies to increase the probability of evasion while maintaining malware functionality. Malware-makeover \cite{lucas2021malware}, MiniMal \cite{li2025minimal}, MalPatch \cite{zhan2023malpatch}, Game-up \cite{labaca-castro2022universal}, \cite{ceschin2019shallow}, \cite{ceschin2020no}, and \cite{fleshman2018static} are examples of such approaches. 

All problem-space evasion attacks share the objective of generating functional malware variants that evade detection. Most guarantee format preservation, while fewer ensure full executability and retention of malicious behavior. Recent works increasingly incorporate sandbox or semantic verification to improve realism.

In this study, we evaluated a representative set of problem-space black-box attacks, including AIMED, AIMED-RL, ARMED, GAMMA and its variants, MAB-malware, Game-Up, and MiniMal against multiple static malware detectors with diverse feature representations. This setup enabled a systematic comparison of attack effectiveness and highlighted potential robustness limitations in current machine learning–based malware detection methods.

%% file: 03_related_work.tex
\section{Related Work}
\label{sec:related_work}
As research on evaluating the robustness of malware detectors against evasion attacks matures, including black-box problem-space attacks, we conducted a comprehensive literature survey to assess the current state of the field. Table \ref{tab:papers_models} in the Appendix summarizes our findings. 

Our survey shows that comprehensive evaluation of malware detectors against problem-space evasion attacks remains lacking. We identify four recurring limitations in the existing literature: limited detector diversity, limited comparison across attacks, insufficient evaluation of realistic problem-space black-box attacks, and limited analysis of cross-model transferability and attack complementarity.

{\textit{Limited detector diversity:} Most studies focus on just two model families, MalConv and LGBM, which are used in 21 and 19 of the 33 studies, respectively. In comparison, other and more recent Windows malware detectors receive much less attention. A similar pattern can be seen in the experimental setups: most studies test only one type of model and one attack. Only two studies evaluate more than three model types, while 12 consider multiple attacks. Commercial detectors are somewhat better represented, appearing in 14 studies, but there is still limited broad and systematic testing across different models and attack strategies. These findings suggest a need for more diverse evaluations that include modern malware detectors and examine their robustness against a wider range of evasion attacks.

\textit{Limited comparison across attacks:} A small number of works have systematically compared existing methods. \cite{imran2024evaluating} evaluates Extend, Full DOS, Shift, FGSM padding plus slack, and GAMMA against MalConv and LGBM under both white-box and black-box settings, and additionally examines attack transferability; however, its experimental scope is relatively narrow, with limited model diversity, restricted datasets, and with limited independent functional validation of the generated adversarial binaries. \cite{louthanova2024comparison} instead compares a different set of generators, including Partial DOS, Full DOS, GAMMA (padding and section-injection variants), and Gym-malware, evaluated against selected antivirus products rather than open-source detectors, and shows that combining generators can improve evasion; however, the evaluation is centered on commercial AV products and a relatively small set of generators, limiting architectural analysis of the underlying detectors. \cite{demetrio2021adversarial} evaluates white-box and black-box attacks against MalConv, two DNN variants, and a GBDT model, explicitly scoping its analysis to a small set of detectors and, while proposing potential mitigations, does not evaluate its attacks against any adversarially hardened models. More recently, \cite{ponte2026exe} proposes a broader benchmark spanning performance, temporal drift, adversarial robustness, and computational overhead across a wider range of detectors, including transfer attacks between models; however, EXE-Bench uses essentially three manipulation families in its adversarial evaluation: FullDOS, content-shift, and GAMMA. Building on these contributions, there remains an opportunity to broaden coverage to more recently proposed attack methods, incorporate defensive strategies, and evaluate a wider and more consistent range of detection models.

\textit{Limited evaluation of realistic problem-space black-box attacks:} Prior studies \cite{doan2023feature,al2018adversarial,lobascio2025adversarial} have primarily focused on feature-space training, using static representations extracted from PE files (e.g., LIEF features) and crafting adversarial examples with gradient-based white-box methods such as FGSM, PGD, or DeepFool. Other works \cite{lucas2023adversarial} have targeted raw-binary classifiers like MalConv and AvastNet, employing scalable adversarial generation pipelines to train models against in-place randomization, displacement, or gradient-guided attacks, showing that even lower-effort adversarial samples can improve robustness across multiple attack strategies. While these studies provide important insights into adversarial robustness, feature-space or white-box attacks do not necessarily capture the constraints faced by an attacker who must produce a valid, functionality-preserving executable without access to the detector internals. \cite{he2025security} has highlighted the limitations of current ML-based malware detection systems and the need for more realistic adversarial evaluation.

\textit{Limited understanding of transferability and attack complementarity:} The transferability of adversarial malware across models has also been studied \cite{suciu2019exploring}, suggesting that adversarial examples often fail to generalize across detectors, contrasting with image classification. However, these analyses typically involve single-step attacks or small datasets, leaving open questions about the cross-model transferability of iterative, functionality-preserving attacks. Moreover, existing studies typically focus on single attacks in isolation, against singular models, leaving the interplay and complementarity of multiple attacks and models largely unexplored.

To address these limitations, we present a unified large scale evaluation that enables systematic cross-attack and cross-model comparison. More broadly, we conduct a holistic evaluation of diverse attack strategies against malware detectors spanning both classical and deep learning representations.

%% file: 05_rqs.tex
\section{Research Questions}
\label{sec:rqs}

We next define the research questions that structure our analysis. 


\begin{center}
    \textit{\textbf{\hypertarget{RQ1}{RQ1}:} What are the key factors that influence the efficacy of state-of-the-art problem-space evasion attacks against diverse Windows malware classifiers?}
\end{center}

Our first question challenges established results extracted from narrow experimental settings (as detailed in Section \ref{sec:related_work}) and aims to identify which factors are most critical for evasion across a diverse ecosystem of classifiers. To address this question, we evaluate the Attack Success Rate (ASR, of nine recent problem-space evasion attacks applied to the test sets of eight representative static malware classifiers, including seven open-source models and one commercial detector. This diverse testbed allows us to directly investigate RQ1 and identify the key factors that truly influence attack efficacy.

\begin{center}
    \textit{\textbf{\hypertarget{RQ2}{RQ2}:} What is the optimal ensemble of malware evasion attacks?}
\end{center}

Having shown the relative effectiveness (or lack thereof) of established attacks in fooling diverse models, we raise the question of whether an attacker could combine multiple attacks into an ensemble and get significant improvement in success rate. This strategy was used in the computer vision domain to form what is today the strongest baseline for adversarial attacks on images, i.e. AutoAttack~\cite{Croce2020}. In this scenario, an ideal ensemble would be able to produce (almost) all adversarial examples that individual attacks can produce taken together, thereby minimizing the value of integrating the excluded attacks.

To achieve this, we study the overlap and complementarity of individual state-of-the-art attacks, in terms of which detected original malware they can turn into a successful adversarial example. More precisely, for an attack $A$ and any other attack $B$, we want to determine how many unique adversarial examples $A$ can produce, i.e., how many of the original malware that $A$ can turn into successful adversarial examples that $B$ cannot. Attacks with a sufficient number of unique malware samples are added to the ensemble.


\begin{center}
\textit{\textbf{\hypertarget{RQ3}{RQ3}:} Do malware evasion attacks transfer across malware detectors?}     
\end{center}

Transferability is an important property of adversarial examples that is practically relevant for malware detection~\cite{nasr2025evaluating}. In real-world settings, the attacker may not know any information about the target model or they have only a limited number (even a single) of query attempts to the target model, e.g., due to query costs or to keep the attack elusive. Transfer attacks are relevant in such scenarios because they generate adversarial malware using a surrogate model (built by the attacker), with no extra cost and no risk of alerting the target system. Furthermore, studying transferability enables the risk estimation of generalizable adversarial malware, able to bypass detection from multiple models and thus to infect a large number of systems.

We, therefore, study the transferability of malware evasion attacks across models.
Starting from a given malware sample, a surrogate model, and an evasion attack (all assumed to be owned and controlled by the attacker), we produce an adversarial example that successfully bypasses the surrogate model. We then test whether this example transfers to a given target model (assumed unknown to the attacker). We repeat this process across all combination of (surrogate and target) models, attacks and original samples. Through this extensive study, we aim to uncover whether attacks remain as effective in a transfer-based threat model as they were in direct attack settings, and which attacks have the highest transferability -- highlighting the generalization ability of the examples they produce.

\begin{center}
    \textit{\textbf{\hypertarget{RQ4}{RQ4}:} How effective is adversarial hardening in improving robustness in single-attack and cross-attack scenarios? 
    }
\end{center}

We next investigate the effectiveness of defenses. In other domains, adversarial hardening has been recognized as the only defense that stood the test of time~\cite{madry2017towards, yu2019interpreting} and was shown to be effective in the malware domain as well\cite{lucas2023adversarial}. We thus naturally focus on this well-established mechanism.

Accordingly, for a given model $m$ and attack $a$, we use $a$ generate adversarial examples from the training set of $a$. We then retrain $m$ from scratch using the produced adversarial examples, with the same training parameters as the original model. We then attack each model with the same protocol as in RQ1 and compute the resulting success rates.

Moreover, the fact that new problem-space attacks continually emerge raise the question of whether a model hardened against a given attack is also robust to different attacks, or if it ``overfits'' to the specific attack patterns it learned from. We, therefore, evaluate model robustness across attacks by hardening a given model $m$ against a given attack $a$, and then evaluating the robustness of the hardened model to different attacks and compare it to the vanilla model's. Thereby, we aim to reveal whether (or not) hardening can fundamentally strengthen the model’s understanding of fundamental malware representations, or if it merely mitigates specific obfuscation techniques.

By investigating these research questions, we aim to advance the understanding of requirements for reliable robustness evaluation of malware classifiers, with a specific focus on the influential roles of evasion attacks and defense mechanisms. We translate the insights gained into concrete recommendations for the research community, thereby fostering more rigorous and profound evaluation practices for malware classifiers.

%% file: 06_exp_setting.tex
\section{Evaluation Framework and Settings}
\label{sec:exp_setup}

This study conducts a comprehensive and realistic evaluation of malware classifiers under problem-space evasion attacks using a framework that separates the experimental factors, adversarial setting, and evaluation outcomes. The experimental factors comprise \textit{data}, \textit{representation}, \textit{model}, and \textit{attack}. The \textit{data} factor captures dataset provenance, composition, and scale, which can influence model generalization and robustness. The \textit{representation} factor describes how malware is presented to the detector, ranging from raw binaries and images to hand-crafted PE features. The \textit{model} factor captures differences in detector architectures and learning paradigms. The \textit{attack} factor characterizes how adversarial malware is generated, including the problem-space transformations, search strategy, and functional constraints required to preserve malware validity and behavior \cite{pierazzi2020intriguing,bostani2024evadedroid}.

\begin{figure}[htbp]
    \centering
    \includegraphics[width=0.9\linewidth,keepaspectratio]{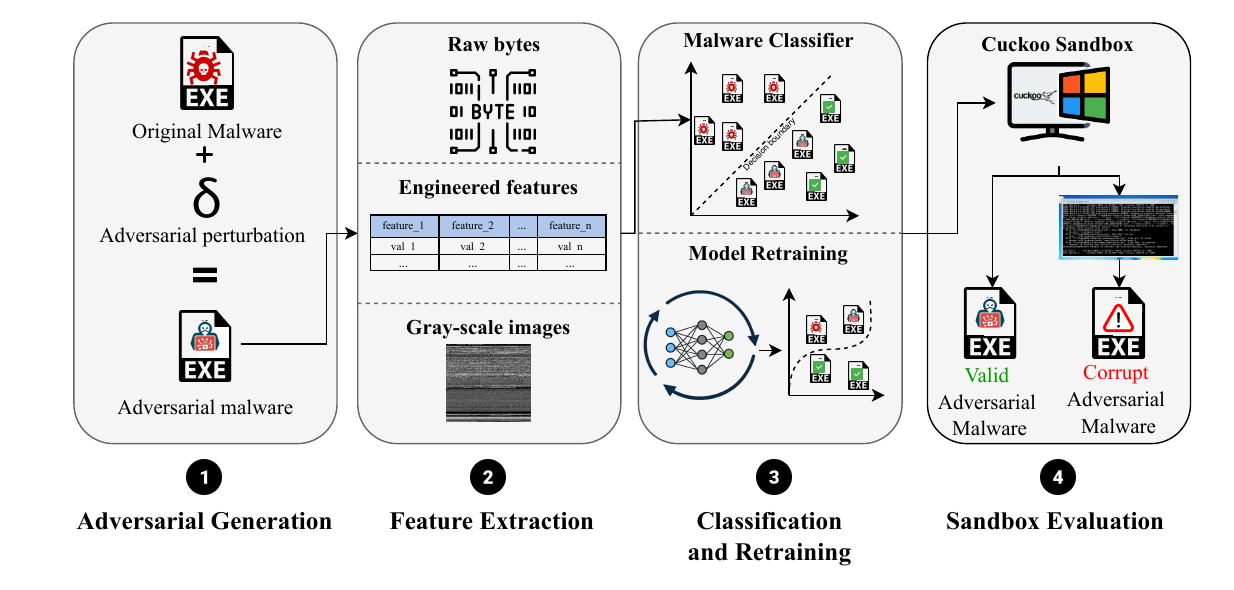}
 \caption{\centering Overview of the adversarial malware generation, evaluation, and retraining pipeline.}
 \label{fig:framework}
\end{figure}

These experimental factors are considered under a \textit{black-box threat model}, which defines the adversary's knowledge and capabilities. Specifically, the adversary has no access to the detector's internal architecture, parameters, or training data and can only query the detector to observe its outputs. This setting reflects a realistic attack scenario while enabling executability-preserving adversarial malware generation through problem-space modifications \cite{he2023efficient,li2023black}. On this foundation, we evaluate robustness from three complementary perspectives: \textit{direct attack effectiveness}, \textit{transferability}, and \textit{adversarial hardening}. Direct attack effectiveness measures evasion against the target detector, transferability captures cross-detector evasion, and adversarial hardening assesses whether adversarial exposure improves robustness to later attacks. Together, these dimensions enable systematic analysis of the interplay among data, representations, models, and attacks across evaluation settings, as illustrated in Figure~\ref{fig:framework}.

\subsection{Models}
\label{subsec:models}

Our evaluation includes eight pre-trained models. The models were selected to cover a diverse set of feature representations and architectures that are widely used. These models are highly relevant because they provide pretrained publicly available artifacts that are routinely used as baselines in adversarial malware research, allowing fair comparison and reproducibility.

\textbf{EMBER GBDT 2017} \cite{anderson2018ember} is a gradient-boosted decision tree trained on engineered PE features, called ember features (v1), containing 2351 features, including section metadata, imports, and byte histograms. 

\textbf{EMBER GBDT 2018} \cite{anderson2018ember} is an updated version of the EMBER GBDT model, trained on a newer dataset distribution, ember feature v2 containing 2381 features, while keeping the same model structure. 

\textbf{EMBER GBDT 2024} \cite{joyce2025ember2024} relies on  EMBER feature version 3 ("thrember"), which enriches previous versions of EMBER with new features (e.g Authenticode signatures, warnings, etc.), increasing the total feature vector dimension from 2,381 to 2,568. We reuse their "All PE files" classifier for our evaluations, which was trained using the Win32, Win64, and .NET files in the training set. 

\textbf{MalConv} \cite{raff2017malware} processes executable binaries as sequences of raw bytes using a 1D convolutional neural network. In our evaluations, we used an alternative version of MalConv that we trained on a subset of files (100K binaries) from the original training set while keeping the architecture unchanged. This approach is adopted because the complete training set used for MalConv (i.e. EMBER 2017) is not publicly available, while our adversarial retraining requires mixing clean examples with adversarial examples crafted from them. Therefore, we have trained this version of MalConv using the available subset of EMBER 2017. 

\textbf{MalConv2} \cite{raff2021classifying} is an improved version of MalConv, designed to handle the extreme length of executable files efficiently. Like MalConv, it operates directly on raw byte sequences, with architectural modifications to address limitations of the original model.  

\textbf{FFNN SOREL} \cite{harang2020sorel} is a feed-forward neural network trained on engineered PE features for malware classification. Multiple random seeds were evaluated, with the best-performing model (seed 3) selected for experiments. 

\textbf{LGBM SOREL} \cite{harang2020sorel} is a LightGBM model trained on engineered PE features for malware detection. The best-performing model (seed 0) among different seeds was selected.

\textbf{Commercial} is a closed source, commercial malware detector that operates on grayscale image representations of executable files. It uses a CNN-based architecture fine-tuned for malware detection.

All models are pre-trained on standard malware datasets: EMBER GBDT 2017 on EMBER 2017, MalConv2 and EMBER GBDT 2018 on EMBER 2018, FFNN SOREL and LGBM SOREL on SOREL-20M, EMBER GBDT 2024 on EMBER 2024, the commercial image-based model on its proprietary dataset, and MalConv on a subset of files from EMBER 2017 dataset. We report the evaluation metrics of the model on clean samples in Table~\ref{tab:model_clean_performance} in Appendix \ref{clean_performance_of_models}. Following established practice in malware detection, we calibrate each detector's decision threshold at a fixed low false-positive operating point rather than using an arbitrary default threshold. We initially target an FPR of 0.1\%, consistent with prior static malware detection studies emphasizing deployability under stringent false-positive constraints\cite{saxe2015deep},\cite{harang2020sorel}. Both 0.1\% and 1\% FPR are commonly used low-FPR operating points in malware-classifier evaluation, with the former representing a more conservative deployment regime and the latter providing a less restrictive but still operationally relevant trade-off \cite{anderson2018ember},\cite{joyce2025ember2024}. For models whose TPR falls below 70\% at 0.1\% FPR, we therefore relax the operating point to 1\% FPR. Consequently, every evaluated detector achieves at least 70\% TPR at its calibrated operating point.




\subsection{Attacks}
\label{subsec:attacks}

The models described in Section \ref{subsec:models} were evaluated against nine black-box problem-space adversarial attacks:

\textbf{MAB-malware} \cite{song2020mab} is a reinforcement learning framework that models the adversarial process as a multi-armed bandit, balancing exploration and exploitation to identify effective transformations for evading malware detectors.

\textbf{Gamma-shift, Gamma-padding, Gamma-section.} 3 variations of the GAMMA attack\cite{demetrio2021functionality,demetrio2021secmlmalware}  based on a genetic algorithm that applies functionality-preserving manipulations, including padding, section injection, and shifting, to generate adversarial malware.

\textbf{Game-up} \cite{labaca-castro2022universal} constructs chains of problem-space transformations using a greedy search strategy to maximize a universal evasion rate across malware samples, effectively flipping their labels to benign.

\textbf{ARMED} \cite{labaca-castro2019armed} applies random binary-level modifications to malware, followed by functionality testing in a sandbox, iterating until evasion is achieved.

\textbf{AIMED} \cite{labaca-castro2019aimed} uses a genetic algorithm to generate adversarial malware by evaluating perturbations for functionality, evasion success, similarity to the original, and diversity, with the fittest samples used to produce subsequent generations. 

\textbf{AIMED-RL} \cite{labaca-castro2021aimed-rl} is a reinforcement-learning based attack using a DQN agent to learn effective transformations, with mechanisms to encourage diversity and achieve high evasion rates with fewer modifications. 

\textbf{MiniMal} \cite{li2025minimal}: MiniMal is a black-box hard-label attack that performs hierarchical perturbation minimization via action reduction, binary search, and particle swarm optimization under functionality constraints.

To select representative attacks, we conducted a thorough literature review and established a set of selection criteria: the selected attacks operate in the problem space; assume black-box access to the target model (or can be adapted from white-box settings); and provide open-source artifacts such as pseudo-code or code implementations. Additionally, we prioritized attacks that are well studied and considered state-of-the-art in the literature.

The resulting set covers widely studied and practically relevant families of problem-space malware evasion techniques, including reinforcement learning methods (MAB-malware, AIMED-RL), evolutionary and genetic algorithms (AIMED, GAMMA), greedy search (Game-up, MiniMal), and random mutation strategies (ARMED). These attacks operate through a commonly used set of functionality-preserving transformations for Windows executable files, detailed in Appendix~\ref{subsec:techniques}, and collectively capture the dominant strategies used in recent adversarial malware research.

Other attacks identified in the literature were excluded because they relied on the same transformation sets or search algorithms as the selected methods, and therefore would not add meaningful diversity to the evaluation. This overlap arises because many attacks build upon the Gym-Malware environment \cite{endgameinc_2017}. Some attacks were excluded due to the lack of reproducibility, as discussed in Sec. \ref{sec:limitations}. 

To ensure a fair comparison, all attacks were executed under comparable effort constraints. While different evasion methods encode the notion of configurations specific to the techniques they employ (e.g., the penalty term in GAMMA or the bandit related parameters in the MAB-Malware attack) the black-box settings common to all attacks are kept identical. Specifically, the query budget is set to 50, the transformation budget to 10, 5 optimization rounds, and all attacks operate in a hard-label setting. Each attack also involves additional method-specific parameters, which are configured according to the recommendations of their respective authors. 


\subsection{Datasets} 
\label{subsec:datasets}
\textbf{Used for robustness evaluation.} To evaluate model robustness, we curated a set of 1,000 malware samples from a larger test pool of approximately 65k PE files collected from VirusShare 499 \cite{virusshare.com_2020}, the Dike Dataset \cite{iosif_2022}, and the commercial test set (files collected from MalwareBazaar \cite{malwarebazaar_malware_sample_exchange} and other sources). All samples in this pool were strictly disjoint from the training data used for any model and are valid functional PE files. From this pool, we sampled 1,000 files using the STAS sampling technique \cite{chow2025breaking},  consistently classified as malware by all evaluated models. STAS combines statistically representative sampling with domain-specific constraints to reduce sampling bias and preserve diversity along multiple axes. We adapt this principle to our PE malware pool so that the resulting subset is not dominated by a narrow group of temporally similar samples. This set will be referred as \textbf{Unified Evaluation Set (UES)}. 

To ensure that our insights are general and not artifacts of dataset selection, we also performed additional experiments across datasets by sampling 1,000 malware instances from the test set of the dataset originally used for each model, which were correctly classified as malicious by the corresponding model. We refer to these as the \textbf{Native Test Set (NTS)}.  These results are included in the Appendix \ref{subsec:attack_effectiveness}.

\textbf{Used for adversarial hardening.} 
To perform adversarial hardening, we randomly selected 15\% malware of the training set corresponding to each model. Adversarial examples were then generated from these samples using the attacks described in Section \ref{subsec:attacks}, and used to improve the robustness of the models by retraining. Additional details about the datasets are provided in Appendix~\ref{subsec:datasets_appendix}. 

\subsection{Sandboxing environment and executability checks}
\label{subsec:sandboxing}

Our study recognizes the importance of ensuring that adversarial samples fooling detection are actually executable. We validate the executability of every adversarial sample to avoid overstating attack success. Although previous studies have recognized this need, they evaluate it only through spot checks on a limited set of examples (10 or 50) \cite{song2020mab,lucas2021malware,zhan2023malpatch}; in contrast, we assess every generated candidate. A sample is counted as a successful evasion only if it both evades the detector and executes successfully.

We validate all adversarial samples using the Cuckoo sandbox \cite{cuckoo_2024}. Several attacks already incorporate executability checking during generation: FAME-based \cite{labaca-castro2023fame} attacks (AIMED, AIMED-RL, ARMED, Game-up)  natively filter broken samples as part of their attack pipeline. For MiniMal, we retain its static file checking, but replace the original functionality checking pipeline with Cuckoo Sandbox to ensure consistency across all evaluated attacks. This modification is necessary for two reasons: (1) to maintain a uniform executability evaluation framework across all methods, and (2) because the original MiniMal implementation relies on proprietary IDA Pro \cite{hex-rays_2024}-based analysis. The remaining attacks do not perform executability checking as part of their attack procedures; for these attacks, we post-validate the generated samples using Cuckoo Sandbox. Samples that execute without crashing are labeled \emph{executable evasive samples}, whereas crashes or sandbox failures are labeled \emph{non-executable} and excluded from subsequent analyses.

We note that executability alone does not guarantee complete preservation of malicious behavior; accordingly, our executability check is intended as a validation of operational validity, rather than a direct measurement of behavioral equivalence with the original malware.


\subsection{Evaluation Metrics}
\label{subsec:eval_metrics}

\textbf{ASR.} The effectiveness of each attack in \textbf{RQ1} is measured using the Attack Success Rate (ASR), which quantifies the proportion of originally detected malware samples that are successfully modified to evade detection while remaining executable. Formally, for a set of malware samples $\mathcal{M}$ and an attack $a$, let $m' \in \mathcal{M}'$ denote the executable modified version of $m \in \mathcal{M}$ generated by $a$. Then,

\[
\mathrm{ASR}(a) = \frac{
\left| \left\{ m \in \mathcal{M} \mid m' \in \mathcal{M}' \text{ evades detection} \right\} \right| }{|\mathcal{M}|}
\]

\textbf{Coverage (C).} Coverage is used in \textbf{RQ2} to compare attacks and determine how much the evasions of one attack overlap with those of another. Let $\mathcal{E}_{\mathrm{A}}$ denote the set of executable malware samples evaded by the most successful attack on average (across models), or by an ensemble of attacks, and $\mathcal{E}_{\mathrm{B}}$ the set of executable malware samples evaded by another attack. The set of overlapping samples is $\mathcal{O} = \mathcal{E}_{\mathrm{A}} \cap \mathcal{E}_{\mathrm{B}}$. The coverage of the $B$ attack by the single or ensemble attacks $A$, $C$, is then: 
\[
C(\mathcal{E}_{\mathrm{A}}, \mathcal{E}_{\mathrm{B}}) = \frac{|\mathcal{O}|}{|\mathcal{E}_{\mathrm{B}}|}
\]
This metric captures the fraction of samples evaded by a given attack that are also evaded by the another successful attack or ensemble of attacks on average, helping to establish a hierarchy of attack effectiveness.

\textbf{Transferability (TR).} Transferability is relevant to \textbf{RQ3}, quantifying how adversarial examples generated for one model generalize to another. For a given attack, we denote by $TR_{A \Rightarrow B}$ its \emph{transferability rate} from the surrogate model $A$ to the target model $B$, i.e., the proportion of executable adversarial samples $m'_A$ successfully produced by the attack on $A$ that also successfully evade detection by model $B$. We also denote by $ASR^{\text{sur}}_{A \Rightarrow B}$ the attack success rate in transfer-based settings, i.e. the proportion of eligible original samples whose adversarial variants evade both $A$ and $B$. 

\[
TR_{A \Rightarrow B} = \frac{ \left| \left\{ m'_A \in \mathcal{M}'_A \mid m'_A \text{ evades } B \right\}\right|}{\left|\mathcal{M}'_A\right|}
\]

\[
ASR^{\mathrm{sur}}_{A \Rightarrow B} = \frac{ \left| \left\{ m \in \mathcal{M} \mid m'_A \text{ evades } A \land m'_A \text{ evades } B \right\} \right| }{ \left|\mathcal{M}\right|}
\]

$TR_{A \Rightarrow B}$ is conditioned on successful evasion of the surrogate, whereas $ASR^{\text{sur}}_{A \Rightarrow B}$ measures end-to-end transfer success over all eligible samples. 


\textbf{Delta ASR ($\Delta$ASR).} To evaluate the impact of adversarial hardening and the transferability of defenses across attacks (\textbf{RQ4}), we measure the change in ASR before and after adversarial retraining. Let $\mathrm{ASR}_{\mathrm{before}}$ denote the ASR of a model prior to hardening, and $\mathrm{ASR}_{\mathrm{after}}$ the ASR after adversarial retraining. We define two variants of $\Delta$ASR, the absolute and relative change, defined as:

\[
\Delta \mathrm{ASR}_{\mathrm{abs}}
=
\mathrm{ASR}_{\mathrm{after}}
-
\mathrm{ASR}_{\mathrm{before}}, 
\quad
\Delta \mathrm{ASR}_{\mathrm{rel}}
=
\frac{
\Delta \mathrm{ASR}_{\mathrm{abs}}
}{
\mathrm{ASR}_{\mathrm{before}}
}
\]

%% file: rq1.tex
\subsection*{\textbf{\hyperlink{RQ1}{RQ1}: Attack Effectiveness}}
Table \ref{tab:attack_perf} reports the ASR of nine black-box, problem-space adversarial attacks evaluated against eight Windows malware detectors on the Unified Evaluation Set (UES) dataset. Due to the stochastic nature of the attacks, each experiment is repeated three times, and we report the mean ASR across the three runs while the standard deviation remains below 1 percentage point in all cases. The results are stable across repetitions and are not driven by individual random runs, which supports the reproducibility of the observed attack and robustness trends.

For completeness, the Appendix includes an additional Table (Tab. \ref{tab:asr_with_func}) comparing ASR values before and after executability-preservation checks for the GAMMA attack variants MAB-malware attack and the MiniMal attack, since such checks are not an integrated component of the attacks themselves but are applied post-hoc to ensure their validity. Moreover, in Appendix \ref{subsec:attack_effectiveness} (Table \ref{tab:combined_asr}), we provide the ASR results for the models evaluated on the Native Test Set (NTS) dataset. The trends of the attacks remain largely consistent across both evaluation settings, yielding the same effective ranking of the attacks.

\begin{table*}[ht!]
\centering
\small
\setlength{\tabcolsep}{3pt} 
\caption{ASR against Windows malware detectors, reported as Mean over 3 runs on the UES dataset. \textbf{Bold} entries indicate the highest mean ASR for each attack, \underline{underlined} entries indicate the second-highest mean ASR.} 
\label{tab:attack_perf}
\resizebox{\textwidth}{!}{%
\begin{tabular}{lccccccccc|c}
\toprule

\textbf{Model} & 
\makecell{\textbf{MAB-}\\\textbf{malware}} & 
\makecell{\textbf{Gamma-}\\\textbf{shift}} & 
\makecell{\textbf{Gamma-}\\\textbf{padding}} & 
\makecell{\textbf{Gamma-}\\\textbf{section}} & 
\makecell{\textbf{MiniMal}} & 
\makecell{\textbf{AIMED}} & 
\makecell{\textbf{AIMED-}\\\textbf{RL}} & 
\makecell{\textbf{ARMED}} & 
\makecell{\textbf{Game-}\\\textbf{up}} & 
\textbf{Avg} \\

\midrule
FFNN-SOREL
& \textbf{85.63}
& 15.97
& 14.17
& 3.27
& \underline{65.20}
& 5.05
& 0.57
& 0.60
& 0
& 21.16 \\

LGBM-SOREL
& \textbf{95.30}
& 70.50
& 70.40
& 3.93
& \underline{81.50}
& 16.23
& 7.77
& 5.20
& 4.40
& 39.47 \\

EMBER-17
& \textbf{89.03}
& 48.83
& 48.87
& 5.43
& \underline{64.80}
& 0.93
& 0.20
& 0.13
& 0.30
& 28.72 \\

EMBER-18
& \textbf{98.33}
& \underline{91.93}
& 91.00
& 4.60
& 77.40
& 56.30
& 30.53
& 19.50
& 26.00
& 55.07 \\

EMBER-24
& \textbf{83.40}
& 12.73
& 7.33
& 0
& \underline{30.60}
& 1.77
& 0.03
& 0.23
& 0
& 15.12 \\

MalConv
& \underline{98.83}
& \textbf{100.00}
& 98.53
& 4.53
& 98.20
& 22.40
& 14.30
& 13.60
& 12.50
& 51.43 \\

MalConv2
& \textbf{98.70}
& \underline{88.53}
& 81.97
& 2.20
& 11.00
& 35.15
& 22.27
& 19.23
& 25.80
& 42.76 \\

Commercial
& \textbf{99.90}
& \underline{97.80}
& 91.13
& 3.20
& 97.23
& 15.70
& 7.70
& 8.30
& 9.50
& 47.83 \\

\midrule

Avg
& \textbf{93.64}
& \underline{65.79}
& 62.92
& 3.39
& 65.74
& 19.19
& 10.42
& 8.35
& 9.81
& 37.70 \\

\bottomrule
\end{tabular}%
}
\end{table*}

Among all attacks methods, MAB-Malware is the most effective attack, achieving the highest average ASR (93.64\%) and almost complete evasion on the commercial detector. Gamma-shift and MiniMal also perform strongly, with average ASRs of 65.79\% and 65.74\%, respectively, and all three attacks remain the most effective even against the newer EMBER-24 model, which otherwise shows more robustness against other attacks.

In our experiments, MAB-Malware operated over seven PE transformations, all of which are contained in the ten-action FAME transformation space; thus, its action space is a strict subset of the FAME action space. Nevertheless, as shown in Table~\ref{tab:attack_perf}, MAB-Malware achieves a substantially higher average ASR (93.64\%) than the FAME-based attacks, whose average ASRs range from 8.35\% to 19.19\%. This advantage is not explained by a larger transformation set or longer perturbation sequences. As shown in Table~\ref{tab:attack_effort}, successful MAB-Malware evasions use only 2.89 transformations on average and 1.55 distinct transformation types, compared with 10.00 and 6.40 for AIMED and 10.00 and 10.00 for ARMED. Although MAB-Malware requires slightly more target-model queries among successful evasions, these query counts are success-conditioned and should therefore be interpreted jointly with ASR.

We further analyze the composition of successful perturbation sequences. MAB-Malware exhibits a concentrated transformation profile: section addition and overlay append occur in 54.16\% and 51.65\% of successful evasions, respectively, and also account for 47.20\% and 40.97\% of the final actions preceding the first observed evasion. In contrast, AIMED distributes successful sequences across a broader portion of its action space. These results show that transformation-set breadth alone does not explain attack effectiveness and suggest that the mechanisms used to search, select, and compose transformations are important contributors to MAB-Malware's higher evasion success.

\begin{table*}[t]
\centering
\caption{Attack effectiveness and effort. Query and perturbation statistics are averaged over successful attacks; ASR is averaged over all models.}
\label{tab:attack_effort}

\resizebox{\textwidth}{!}{
\begin{tabular}{l l c c c c c c}
\toprule
\textbf{Attack} & \textbf{Search strategy} & \textbf{Transforms} &
\textbf{Queries} & \textbf{Seq.\ length} & \textbf{Distinct transforms} &
\textbf{Runtime (s)} & \textbf{ASR (\%)} \\ 
\midrule

\textbf{MAB-Malware}
& Thompson-sampling MAB
& 7
& \textbf{6.51}
& \textbf{2.89}
& \textbf{1.55}
& \textbf{5.17 s}
& \textbf{93.64} \\

AIMED
& Genetic programming
& 10
& 5.79
& 10.00
& 6.40
& 424.60
& 19.19 \\

AIMED-RL
& Deep reinforcement learning
& 10
& 2.13
& 2.13
& 1.93
& 1.95
& 10.42 \\

ARMED
& Random perturbation sequence
& 10
& 1.18
& 10.00
& 10.00
& 14.75
& 8.35 \\

Game-Up
& Universal perturbation search
& 10
& 1.00
& 9.00
& 4.33
& 20.10
& 9.81 \\

\bottomrule
\end{tabular}
}
\end{table*}

Notably, from \ref{tab:asr_with_func}), the Gamma-section attack is heavily penalized by executability-preservation checks: its ASR drops by an average of 94\%, compared with minor decreases of < 1\% for Gamma-padding, MAB-malware and MiniMal, and no decrease for Gamma-shift (see Table \ref{tab:asr_with_func}). This sharp reduction suggests that section-level modifications frequently violate execution constraints. \cite{imran2024evaluating} evaluates the Gamma-sections attack and reports a high number of evasions against MalConv (and transferability to other detectors such as EMBER-17). However, they do not report any runtime execution or functionality checks on the generated adversarial PE files. Our executability validation shows that, although Gamma-sections initially yields many evasions, a substantial fraction of the generated binaries fail to preserve the original malware executability. We provide more details about this behavior in Appendix \ref{subsec:attack_results}. 
 
\textbf{Insights from RQ1:}
The effectiveness of problem-space malware evasion attacks varies substantially across target models, with both ASR and attack ranking changing according to the detector being attacked. Our results indicate that this variability is associated with both the transformation space and the mechanism used to search and compose transformations. MAB-Malware achieves the highest average ASR despite operating over a strict subset of the FAME transformation space, while successful evasions typically rely on only a small number of distinct transformations. Section addition and overlay append dominate successful MAB-Malware sequences, and effective attack construction depends more on identifying and combining high-yield transformations than on simply enlarging the available action set. 

Detector architecture also plays an important role. Raw-byte end-to-end models are generally more susceptible to the evaluated problem-space manipulations, whereas detectors based on engineered PE features are more resistant to several attacks. This highlights the need for functionality-preserving transformations that meaningfully alter the representations used by feature-engineered detectors, rather than only modifying binary structure at a superficial level. More broadly, evaluating attacks across heterogeneous detector families is necessary to avoid conclusions that are specific to a narrow model class or experimental setting. 

%% file: rq2.tex
\subsection*{\textbf{\hyperlink{RQ2}{RQ2}: Attack Complementarity}}

We adopt an iterative coverage-based approach to identify a compact and effective ensemble of attacks. We first evaluate MAB-Malware individually, as it was the strongest single attack in RQ1, and measure how many adversarial examples generated by the remaining attacks are also evaded by MAB-Malware. The corresponding results are provided in Appendix~\ref{subsec:mab_coverage} (Table~\ref{tab:coverage_matrix_mab_reordered}). The observed coverage shows that MAB-Malware does not fully capture the evasive behavior space  observed in our evaluation: a subset of samples evaded by other attacks remains uncovered. We therefore iteratively add Gamma-shift, the second most effective attack from RQ1, which complements MAB-Malware by targeting samples that the latter fails to evade, and evaluate the coverage of the resulting two-attack ensemble.

\begin{table*}[ht!]
\centering
\setlength{\tabcolsep}{5pt} 
\caption{Coverage of other attacks by the ensemble attack (MAB-malware+Gamma-shift) samples across models. Each attack column is divided into two sub-columns:  \texttt{Overlapping / Evasive} and \texttt{Coverage (\%)}.}
\label{tab:coverage_matrix_subcolumns}
\resizebox{\textwidth}{!}{%
\small 
\begin{tabular}{l|ll|ll|ll|ll|ll|ll|ll}
\toprule 
\textbf{Model} 
& \multicolumn{2}{c|}{\makecell{\textbf{Gamma-}\\\textbf{padding}}}
& \multicolumn{2}{c|}{\makecell{\textbf{Gamma-}\\\textbf{section}}}
& \multicolumn{2}{c|}{\textbf{MiniMal}}
& \multicolumn{2}{c|}{\textbf{AIMED}}
& \multicolumn{2}{c|}{\textbf{AIMED-RL}}
& \multicolumn{2}{c|}{\textbf{ARMED}}
& \multicolumn{2}{c}{\makecell{\textbf{Game-}\\\textbf{Up}}} \\

\midrule

FFNN-SOREL
& 142 / 142 & 100\%
& 27 / 30 & 90\%
& 652 / 652 & 100\%
& 50 / 50 & 100\%
& 5 / 5 & 100\%
& 6 / 6 & 100\%
& 0 / 0 & -- \\

LGBM-SOREL
& 704 / 704 & 100\%
& 40 / 40 & 100\%
& 815 / 815 & 100\%
& 162 / 162 & 100\%
& 72 / 76 & 94.74\%
& 52 / 52 & 100\%
& 44 / 44 & 100\% \\

EMBER-17
& 486 / 488 & 99.59\%
& 54 / 54 & 100\%
& 648 / 648 & 100\%
& 9 / 9 & 100\%
& 2 / 2 & 100\%
& 2 / 2 & 100\%
& 3 / 3 & 100\% \\

EMBER-18
& 910 / 910 & 100\%
& 46 / 46 & 100\%
& 774 / 774 & 100\%
& 563 / 563 & 100\%
& 305 / 305 & 100\%
& 195 / 195 & 100\%
& 260 / 260 & 100\% \\

EMBER-24
& 73 / 73 & 100\%
& 0 / 0 & --
& 306 / 306 & 100\%
& 17 / 17  & 100\% 
& 3 / 3 & 100\%
& 3 / 4 & 75\%
& 0 / 0 & -- \\
 
MalConv
& 986 / 986 & 100\%
& 46 / 46 & 100\%
& 982 / 982 & 100\%
& 224 / 224 & 100\%
& 143 / 143 & 100\%
& 136 / 136 & 100\%
& 125 / 125 & 100\% \\

MalConv2
& 820 / 820 & 100\%
& 22 / 22 & 100\%
& 110 / 110 & 100\%
& 351 / 351 & 100\%
& 222 / 222 & 100\%
& 193 / 197 & 97.97\%
& 258 / 258 & 100\% \\

Commercial
& 911 / 911 & 100\%
& 32 / 32 & 100\%
& 973 / 973 & 100\%
& 157 / 157 & 100\%
& 77 / 77 & 100\%
& 83 / 83 & 100\%
& 95 / 95 & 100\% \\

\cmidrule(lr){1-15}

Total
& 5032 / 5034 & 99.96\%
& 267 / 270 & 98.89\%
& 5260 / 5260 & 100\%
& 1533 / 1533 & 100\%
& 829 / 833 & 99.52\%
& 670 / 675 & 99.26\%
& 785 / 785 & 100\% \\

\bottomrule
\end{tabular}
}
\end{table*}

Table~\ref{tab:coverage_matrix_subcolumns} summarizes this analysis by reporting, for each attack considered in RQ1, the number of adversarial samples it generates and the proportion of those samples that are also covered by the MAB-Malware+Gamma-shift ensemble.The ensemble achieves complete or near-complete coverage across most attack-model combinations. Aggregated over the evaluated samples, coverage ranges from 98.89\% for Gamma-section to 100\% for MiniMal, AIMED, and Game-Up, with similarly high values for the remaining attacks.

We obtain similar results on the NTS dataset, with MAB-Malware and Gamma-shift again providing the strongest coverage among the evaluated combinations, and the overall coverage trends remaining consistent, implying that the complementarity observed on the primary dataset is not limited to a single sample set. We emphasize that this ensemble is not intended to be exhaustive or to represent the complete space of possible malware attacks. Rather, it provides a compact attack set that captures the adversarial examples observed in our evaluation and can serve as a practical baseline for robustness assessment.

\textbf{Insights from RQ2:}
MAB-Malware and Gamma-shift form the compact attack ensemble identified in our evaluation that achieves near-complete coverage of the adversarial examples generated by the evaluated attacks. 
Their complementarity shows that high individual ASR does not imply complete coverage: attacks with lower overall success can still expose samples missed by the strongest attack. This supports the use of cumulative coverage, rather than ASR alone, when selecting attacks for robustness evaluation. Their complementary behavior allows the ensemble to capture adversarial examples that are missed by either attack individually, while avoiding the additional cost of evaluating a larger collection of attacks. We therefore propose this two-attack ensemble as a practical baseline for evaluating the robustness of malware classifiers. As new attacks are introduced, the same coverage-based procedure can be reapplied to determine whether they expose previously uncovered vulnerabilities and materially expand the evaluation set, while attacks that provide little additional coverage may be considered redundant for this evaluation objective. This iterative procedure provides a potential foundation for developing a standardized and evolving attack suite for malware robustness evaluation, analogous in spirit to established robustness benchmarks such as RobustBench~\cite{croce2021}.

%% file: rq3.tex
\subsection*{\textbf{\hyperlink{RQ3}{RQ3}: Transferability}}

To quantify transferability, we first take an attack-centric perspective and ask whether adversarial examples that successfully evade a surrogate model also evade a different target model. 
For each attack and target model, and for all successful adversarial samples produced for \textit{RQ1}, we test these samples against the target model and compute transferability rate as the proportion of surrogate-model evasions that also evade the target, as formally defined in Section~\ref{subsec:eval_metrics}.

\begin{figure}[ht!]
     \centering
         \centering
         \includegraphics[width=0.7\linewidth]{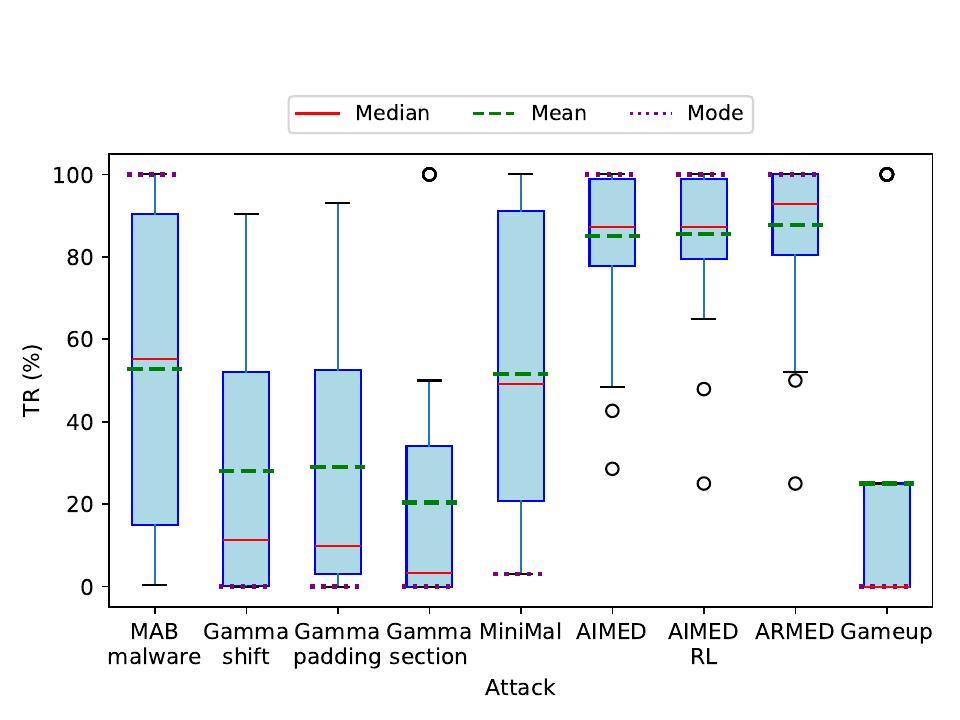}
         \caption{Distribution of average ($TR_{A \Rightarrow B}$ \%) (over all surrogate models) for each attack, over all target models.}
         \label{fig:transferability}

\end{figure}

Figure~\ref{fig:transferability} reports the conditional transferability of successful surrogate-model evasions. AIMED, AIMED-RL, and ARMED exhibit the highest transferability among their successful evasions, while MAB-Malware and MiniMal show more moderate transfer rates. This illustrates that direct attack effectiveness and transferability are distinct properties: attacks with low ASR can nevertheless produce adversarial examples that generalize well across models.

\begin{figure}[ht!]
    \centering

    \begin{subfigure}[b]{0.49\columnwidth}
        \centering
        \includegraphics[width=\linewidth]{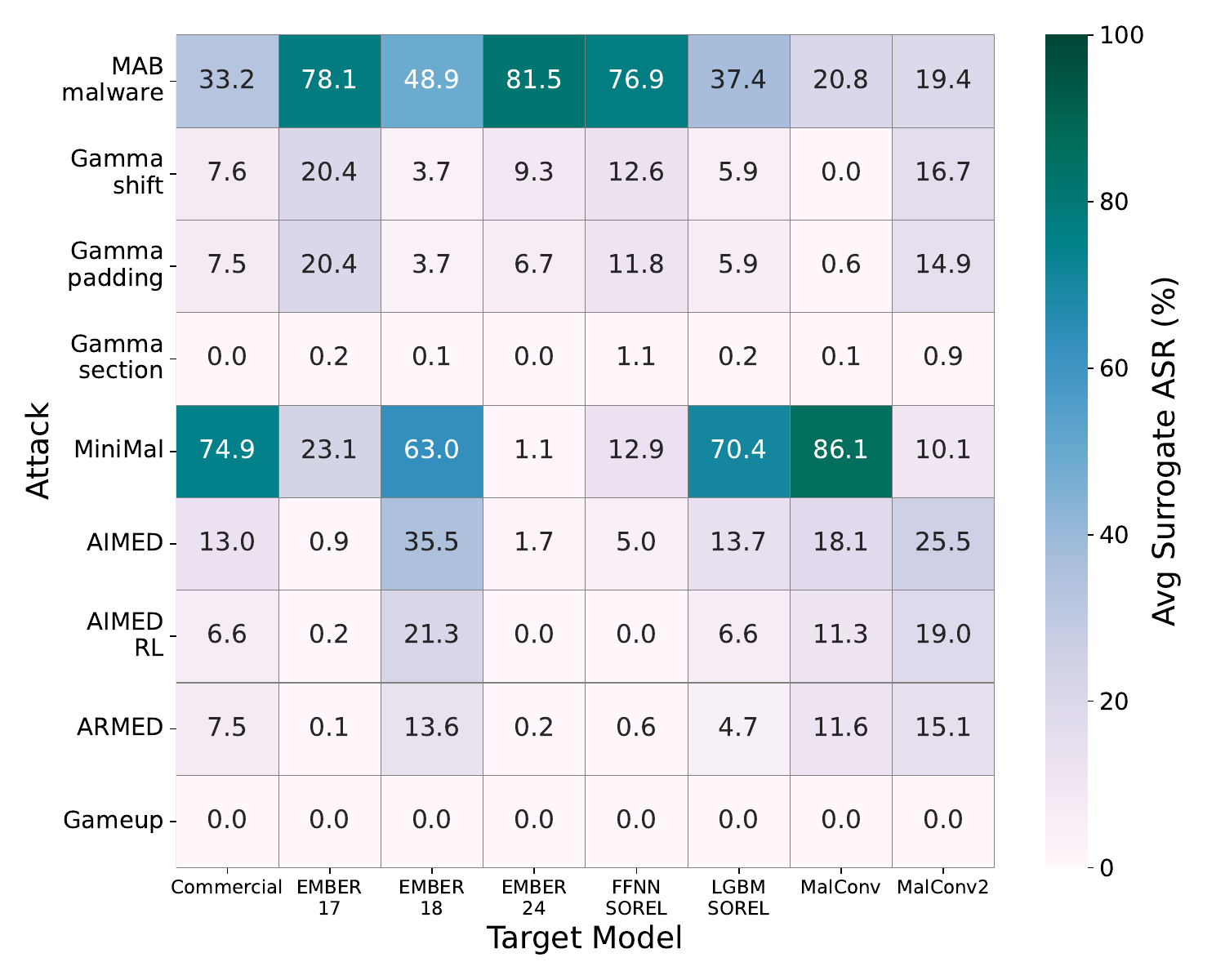}
        \caption{Avg $ASR^{\text{sur}}_{A \Rightarrow B}$ across attack--target model pairs.}
        \label{fig:attack_vs_target_weighted}
    \end{subfigure}
    \hfill
    \begin{subfigure}[b]{0.49\columnwidth}
        \centering
        \includegraphics[width=\linewidth]{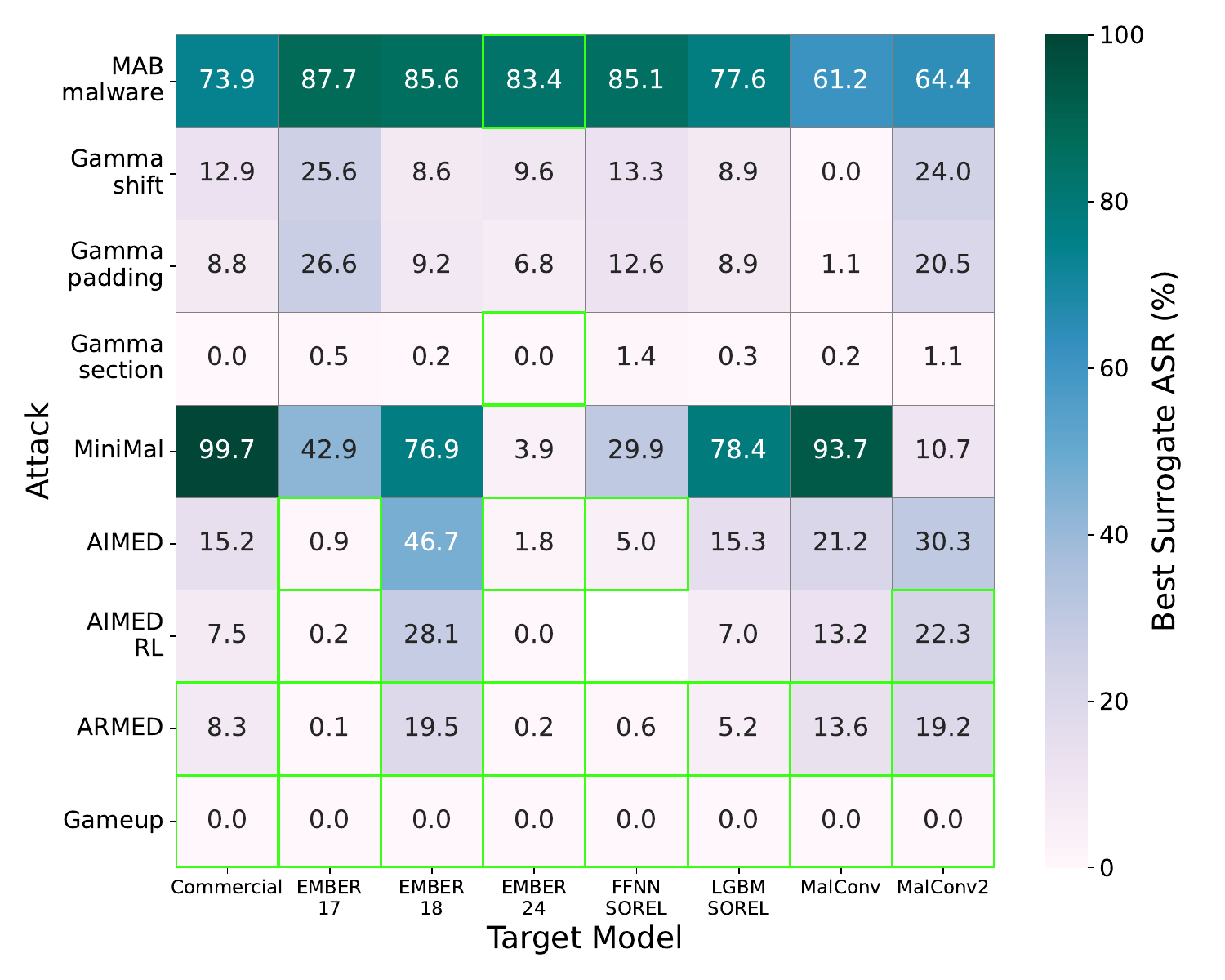}
        \caption{Maximum $ASR^{\text{sur}}_{A \Rightarrow B}$ under an oracle surrogate assumption.}
        \label{fig:max_transferability}
    \end{subfigure}

    \caption{Adversarial attack transferability across models and attacks.}
    \label{fig:combined_transfer}
\end{figure}

We next move from the attack-centric perspective to the target-model perspective. Rather than asking whether successful adversarial examples transfer, we ask: how vulnerable is a given target model when adversarial examples are generated on a different model? We consider two transfer settings: an arbitrary-surrogate setting, where transfer success is averaged over all non-target surrogates, and a best-surrogate setting, where the attacker selects the surrogate yielding the highest transferred ASR.

Figure~\ref{fig:attack_vs_target_weighted} reports the transferred ASR against each target model, averaged over all surrogate models different from the target. Transfer performance is generally lower than direct-attack ASR, but the magnitude of the reduction varies substantially across attacks and targets. MAB-Malware retains comparatively high transferred ASR against several targets, reaching 81.5\% on EMBER-24, 78.1\% on EMBER-17, and 76.9\% on FFNN-SOREL, whereas its average transfer to MalConv and MalConv2 drops to 20.8\% and 19.4\%, respectively. MiniMal exhibits a different pattern, transferring strongly to MalConv (86.1\%), the commercial detector (74.9\%), and LGBM-SOREL (70.4\%), but only weakly to EMBER-24 (1.1\%). These differences demonstrate that transferability depends strongly on the specific attack and surrogate--target model relationship rather than on direct attack effectiveness alone.

Lastly,  Figure~\ref{fig:max_transferability} considers an oracle setting in which the attacker selects the surrogate yielding the highest transferred ASR for each target. Surrogate selection substantially increases transfer success. For MAB-Malware, for example, transferred ASR increases from 33.2\% to 73.9\% against the commercial detector, from 20.8\% to 61.2\% against MalConv, and from 19.4\% to 64.4\% against MalConv2. In some cases, the best-surrogate transfer approaches or matches direct-attack performance, such as MAB-Malware against EMBER-24 (83.4\%). These results show that surrogate choice can substantially alter the apparent transfer robustness of a detector

Notably, transfer can occasionally exceed the corresponding direct-attack ASR. For example, MiniMal reaches 99.7\% transferred ASR against the commercial detector with the best surrogate, compared with 97.23\% under direct attack. This shows that surrogate-generated adversarial examples can expose target-model vulnerabilities that are not necessarily reached by directly optimizing against that target.

\textbf{Insights from RQ3:} Our results confirm that transferability depends strongly on the choice of surrogate and target models, with carefully selected surrogates achieving substantially higher transfer rates than arbitrary ones. However, we also observe an important distinction between \textit{attack effectiveness and transferability}: attacks with lower direct-attack success can nevertheless produce adversarial examples that transfer well, whereas highly effective attacks may produce more surrogate-specific examples. This highlights the importance of evaluating both attack effectiveness and transferability when assessing the threat posed by adversarial malware.

%% file: rq4.tex
\subsection*{\textbf{\hyperlink{RQ4}{RQ4}: Adversarial Hardening}}

\begin{table}[ht!] 
\centering 
\scriptsize 
\setlength{\tabcolsep}{4pt} 
\caption{Absolute and relative change in mean $\Delta \mathrm{ASR}_{\mathrm{abs}}$ and $\Delta \mathrm{ASR}_{\mathrm{rel}}$ before and after adversarial hardening on the UES dataset, illustrating robustness and defense transfer across attacks.} 

\label{tab:robustness_ues_dataset} 
\resizebox{\columnwidth}{!}{

\begin{tabular}{lllcccc} 
\toprule 
\textbf{Model} & 
\textbf{Hardened Against} & 
\textbf{Test Attack} &
$\boldsymbol{\mathrm{ASR}_{\mathrm{before}}}$ & 
$\boldsymbol{\mathrm{ASR}_{\mathrm{after}}}$ &
$\boldsymbol{\Delta \mathrm{ASR}_{\mathrm{abs}}}$ & 
$\boldsymbol{\Delta \mathrm{ASR}_{\mathrm{rel}}}$ (\%) \\ 
\midrule

\multirow{25}{*}{\textbf{EMBER-17}} 
& \multirow{5}{*}{Gamma-padding} 

& \textbf{Gamma-padding} & \textbf{48.87} & \textbf{0.36} 
& \textbf{\good{-48.51}} & \textbf{\good{-99.26\%}} \\ 

& & Gamma-shift & 48.83 & 1.00
& \good{-47.83} & \good{-97.95\%} \\ 

& & MAB-malware & 89.03 & 77.9
& \good{-11.13} & \good{-12.50\%} \\ 

& & MiniMal & 64.80 & 40.5
& \good{-24.30} & \good{-37.50\%} \\ 

& & AIMED & 0.93 & 0
& \good{-0.93} & \good{-100.00\%} \\ 

\cmidrule(lr){2-7} 

& \multirow{5}{*}{Gamma-shift} 

& Gamma-padding & 48.87 & 0.26 
& \good{-48.61} & \good{-99.47\%} \\ 

& & \textbf{Gamma-shift} & \textbf{48.83} & \textbf{0.43} 
& \textbf{\good{-48.40}} & \textbf{\good{-99.12\%}} \\ 

& & MAB-malware & 89.03 & 77.7 
& \good{-11.33} & \good{-12.73\%} \\ 

& & MiniMal & 64.80 & 52.0
& \good{-12.80} & \good{-19.75\%} \\ 

& & AIMED & 0.93 & 0
& \good{-0.93} & \good{-100.00\%} \\ 

\cmidrule(lr){2-7} 

& \multirow{5}{*}{MAB-malware} 

& Gamma-padding & 48.87 & 0.66 
& \good{-48.21} & \good{-98.65\%} \\ 

& & Gamma-shift & 48.83 & 1.4 
& \good{-47.43} & \good{-97.13\%} \\ 

& & \textbf{MAB-malware} & \textbf{89.03} & \textbf{3.6} 
& \textbf{\good{-85.43}} & \textbf{\good{-95.96\%}} \\ 

& & MiniMal & 64.80 & 1.4
& \good{-63.40} & \good{-97.84\%} \\ 

& & AIMED & 0.93 & 0 
& \good{-0.93} & \good{-100.00\%} \\ 

\cmidrule(lr){2-7} 

& \multirow{5}{*}{MiniMal} 

& Gamma-padding & 48.87 & 0.10 
& \good{-48.77} & \good{-99.80\%} \\ 

& & Gamma-shift & 48.83 & 0.80 
& \good{-48.03} & \good{-98.36\%} \\ 

& & MAB-malware & 89.03 & 83.6 
& \good{-5.43} & \good{-6.10\%} \\ 

& & \textbf{MiniMal} & \textbf{64.80} & \textbf{0} 
& \textbf{\good{-64.80}} & \textbf{\good{-100.00\%}} \\ 

& & AIMED & 0.93 & 0 
& \good{-0.93} & \good{-100.00\%} \\ 

\cmidrule(lr){2-7} 

& \multirow{5}{*}{AIMED} 

& Gamma-padding & 48.87 & 83.03 
& \bad{34.16} & \bad{69.90\%} \\ 

& & Gamma-shift & 48.83 & 83.6 
& \bad{34.77} & \bad{71.21\%} \\ 

& & MAB-malware & 89.03 & 85.4 
& \good{-3.63} & \good{-4.08\%} \\ 

& & MiniMal & 64.80 & 60.0 
& \good{-4.80} & \good{-7.41\%} \\ 

& & \textbf{AIMED} & \textbf{0.93} & \textbf{0} 
& \textbf{\good{-0.93}} & \textbf{\good{-100.00\%}} \\  

\cmidrule(lr){1-7} 

\multirow{25}{*}{\textbf{MalConv}} 
& \multirow{5}{*}{Gamma-padding} 

& \textbf{Gamma-padding} & \textbf{98.53} & \textbf{91.54} 
& \textbf{\good{-6.99}} & \textbf{\good{-7.09\%}} \\ 

& & Gamma-shift & 100 & 99.92  
& \good{-0.08} & \good{-0.08\%} \\ 

& & MAB-malware & 98.83 & 98.53 
& \good{-0.30} & \good{-0.30\%} \\ 

& & MiniMal & 98.20 & 82.9 
& \good{-15.30} & \good{-15.58\%} \\ 

& & AIMED & 22.40 & 20.03 
& \good{-2.37} & \good{-10.58\%} \\ 

\cmidrule(lr){2-7} 

& \multirow{5}{*}{Gamma-shift} 

& Gamma-padding & 98.53 & 88.152
& \good{-10.38} & \good{-10.53\%} \\ 

& & \textbf{Gamma-shift} & \textbf{100} & \textbf{99.49} 
& \textbf{\good{-0.51}} & \textbf{\good{-0.51\%}} \\ 

& & MAB-malware & 98.83 & 98.55 
& \good{-0.28} & \good{-0.28\%} \\ 

& & MiniMal & 98.20 & 88.6 
& \good{-9.60} & \good{-9.78\%} \\ 

& & AIMED & 22.40 & 19.63 
& \good{-2.77} & \good{-12.37\%} \\ 

\cmidrule(lr){2-7} 

& \multirow{5}{*}{MAB-malware} 

& Gamma-padding & 98.53 & 89.64 
& \good{-8.89} & \good{-9.02\%} \\ 

& & Gamma-shift & 100 & 99.75 
& \good{-0.25} & \good{-0.25\%} \\ 

& & \textbf{MAB-malware} & \textbf{98.83} & \textbf{97.57} 
& \textbf{\good{-1.26}} & \textbf{\good{-1.27\%}} \\ 

& & MiniMal & 98.20 & 89.8 
& \good{-8.40} & \good{-8.55\%} \\ 

& & AIMED & 22.40 & 17.93 
& \good{-4.47} & \good{-19.96\%} \\ 

\cmidrule(lr){2-7} 

& \multirow{5}{*}{MiniMal} 

& Gamma-padding & 98.53 & 92.45 
& \good{-6.08} & \good{-6.17\%} \\ 

& & Gamma-shift & 100 & 99.65 
& \good{-0.35} & \good{-0.35\%} \\ 

& & MAB-malware & 98.83 & 98.11 
& \good{-0.72} & \good{-0.73\%} \\ 

& & \textbf{MiniMal} & \textbf{98.20} & \textbf{75.7} 
& \textbf{\good{-22.50}} & \textbf{\good{-22.91\%}} \\ 

& & AIMED & 22.40 & 15.8 
& \good{-6.60} & \good{-29.46\%} \\ 

\cmidrule(lr){2-7} 

& \multirow{5}{*}{AIMED} 

& Gamma-padding & 98.53 & 98.50 
& \good{-0.03} & \good{-0.03\%} \\ 

& & Gamma-shift & 100 & 100  
& 0.00 & 0.00\% \\ 

& & MAB-malware & 98.83 & 98.72 
& \good{-0.11} & \good{-0.11\%} \\ 

& & MiniMal & 98.20 & 87.1 
& \good{-11.10} & \good{-11.30\%} \\ 

& & \textbf{AIMED} & \textbf{22.40} & \textbf{8.30} 
& \textbf{\good{-14.10}} & \textbf{\good{-62.95\%}} \\ 

\cmidrule(lr){1-7} 

\multirow{25}{*}{\textbf{Commercial}} 
& \multirow{5}{*}{Gamma-padding} 

& \textbf{Gamma-padding} & \textbf{91.13} & \textbf{0.10} 
& \textbf{\good{-91.03}} & \textbf{\good{-99.89\%}} \\ 

& & Gamma-shift & 97.80 & 1.19 
& \good{-96.61} & \good{-98.78\%} \\ 

& & MAB-malware & 99.90 & 99.90
& 0.00 & 0.00\% \\ 

& & MiniMal & 97.23 & 93.2
& \good{-4.03} & \good{-4.14\%} \\ 

& & AIMED & 15.70 & 20.5
& \bad{4.80} & \bad{30.57\%} \\  

\cmidrule(lr){2-7} 

& \multirow{5}{*}{Gamma-shift} 

& Gamma-padding & 91.13 & 3.18 
& \good{-87.95} & \good{-96.51\%} \\ 

& & \textbf{Gamma-shift} & \textbf{97.80} & \textbf{0.30} 
& \textbf{\good{-97.50}} & \textbf{\good{-99.69\%}} \\ 

& & MAB-malware & 99.90 & 99.90  
& 0.00 & 0.00\% \\ 

& & MiniMal & 97.23 & 88.3 
& \good{-8.93} & \good{-9.18\%} \\ 

& & AIMED & 15.70 & 16.1 
& \bad{0.40} & \bad{2.55\%} \\ 

\cmidrule(lr){2-7} 

& \multirow{5}{*}{MAB-malware} 

& Gamma-padding & 91.13 & 46.31
& \good{-44.82} & \good{-49.18\%} \\ 

& & Gamma-shift & 97.80 & 71.72 
& \good{-26.08} & \good{-26.67\%} \\ 

& & \textbf{MAB-malware} & \textbf{99.90} & \textbf{99.90} 
& \textbf{0.00} & \textbf{0.00\%} \\ 

& & MiniMal & 97.23 & 92.2 
& \good{-5.03} & \good{-5.17\%} \\ 

& & AIMED & 15.70 & 20.3 
& \bad{4.60} & \bad{29.30\%} \\ 

\cmidrule(lr){2-7} 

& \multirow{5}{*}{MiniMal} 

& Gamma-padding & 91.13 & 92.24 
& \bad{1.11} & \bad{1.22\%} \\ 

& & Gamma-shift & 97.80 & 99.69 
& \bad{1.89} & \bad{1.93\%} \\ 

& & MAB-malware & 99.90 & 100 
& \bad{0.10} & \bad{0.10\%} \\ 

& & \textbf{MiniMal} & \textbf{97.23} & \textbf{92.1} 
& \textbf{\good{-5.13}} & \textbf{\good{-5.28\%}} \\ 

& & AIMED & 15.70 & 10.5 
& \good{-5.20} & \good{-33.12\%} \\ 

\cmidrule(lr){2-7} 

& \multirow{5}{*}{AIMED} 

& Gamma-padding & 91.13 & 98.05  
& \bad{6.92} & \bad{7.59\%} \\ 

& & Gamma-shift & 97.80 & 99.45  
& \bad{1.65} & \bad{1.69\%} \\ 

& & MAB-malware & 99.90 & 99.90 
& 0.00 & 0.00\% \\ 

& & MiniMal & 97.23 & 92.7 
& \good{-4.53} & \good{-4.66\%} \\ 

& & \textbf{AIMED} & \textbf{15.70} & \textbf{10.20} 
& \textbf{\good{-5.50}} & \textbf{\good{-35.03\%}} \\  

\bottomrule 
\end{tabular}
}
\end{table}

\begin{figure*}[htbp]
    \centering
    
    \includegraphics[width=\textwidth]{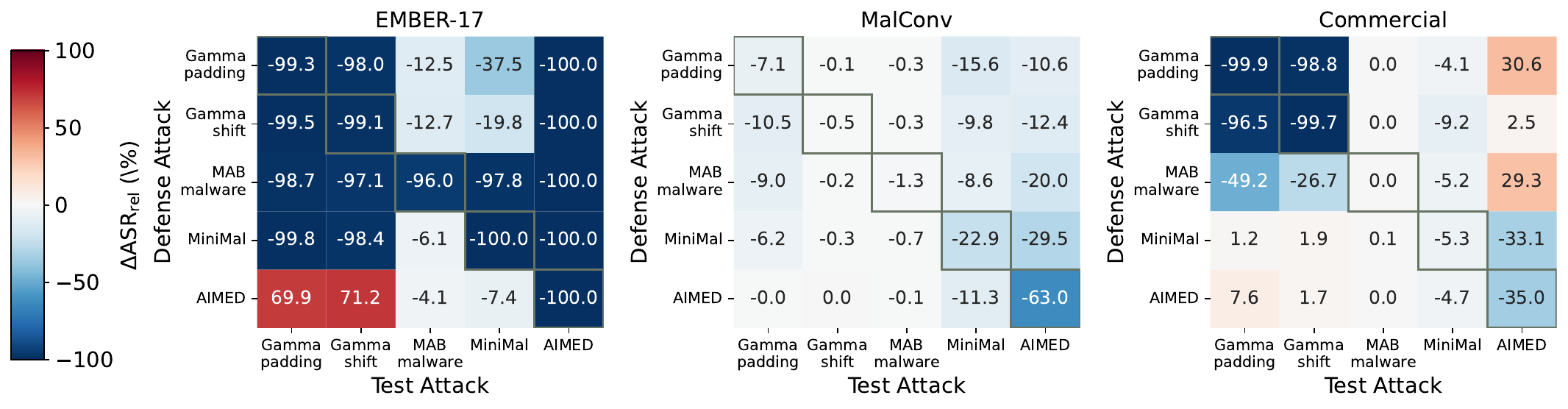}
    \caption{\centering Relative change in ASR ($\Delta \mathrm{ASR}_{\mathrm{rel}}$, \%) after adversarial hardening on the UES dataset. Defense Attack denotes the attack used for training and Test Attack the attack used for evaluation. Negative values indicate reduced ASR (improved robustness), while positive values indicate increased susceptibility.}

    \label{fig:ari_heatmaps}
\end{figure*}

We evaluate adversarial hardening on three representative models, EMBER-17, MalConv, and the Commercial model, which span diverse feature representations and have each shown vulnerability to problem-space attacks. Each model is retrained using adversarial examples from the five attacks with the highest average impact: MAB-malware, Gamma-shift, MiniMal, Gamma-padding, and AIMED, all of which achieve an average ASR above 15\% and as shown in RQ2, MAB-malware and Gamma-shift also achieve broad coverage across other attack types making this set both representative of high-performing attacks.

Table \ref{tab:robustness_ues_dataset} and Figure \ref{fig:ari_heatmaps} present the ASRs of hardened models against both the attack used for hardening and other unseen attacks. Each attack was executed 3 times and results are averaged across runs, whereas the standard deviation remains minimal. For comparison, the corresponding pre-hardening ASRs are also shown, with results for the attack used in hardening highlighted to emphasize direct defense effectiveness. To evaluate the hardened models, we use the same test sets as for the original models, i.e., the UES dataset. Over 90\% of files remain identical across the two evaluations; in the few cases where a file was misclassified, it was replaced with a correctly detected malware sample to maintain consistency. Table \ref{tab:hardened_accuracy} (Appendix \ref{hardened_model_clean_performance}) reports the performance metrics of the adversarially retrained models, with classification thresholds determined using the same procedure as for the clean models (Section \ref{subsec:models}). We additionally report results on the NTS dataset in Appendix \ref{subsec:hardened_model_performance} (Table \ref{tab:robustness}). Notably, AIMED shows a low initial ASR on the EMBER-17 model when evaluated on the UES dataset, but a substantially higher ASR on the NTS dataset; despite this shift in initial ASR, the robustness gains from hardening remain consistent across both settings.

When hardened models are attacked with the same attacks used for retraining, adversarial hardening generally reduces ASR substantially, although the magnitude of the improvement depends strongly on the model. EMBER-17 exhibits the largest and most consistent gains, with relative ASR reductions of 100\% for MiniMal and AIMED, approximately 99\% for Gamma-padding and Gamma-shift, and 95.96\% for MAB-malware. This indicates that the feature-based EMBER-17 detector can effectively adapt to the adversarial patterns represented during retraining. In contrast, MalConv benefits substantially less from attack-specific hardening: the relative reductions are only 7.09\%, 0.51\%, and 1.27\% for Gamma-padding, Gamma-shift, and MAB-malware, respectively, although larger improvements are observed for MiniMal (22.91\%) and particularly AIMED (62.95\%). The Commercial model shows strong robustness gains against Gamma-padding and Gamma-shift, with relative ASR reductions of 99.89\% and 99.69\%, respectively, and a moderate improvement against AIMED (35.03\%); however, hardening against MAB-malware produces no reduction in ASR.

To assess defense transferability, we evaluate each hardened model against attacks not used during retraining. The strongest reciprocal transfer occurs between Gamma-padding and Gamma-shift: on EMBER-17, hardening against Gamma-padding reduces Gamma-shift ASR by 97.95\%, while Gamma-shift hardening reduces Gamma-padding ASR by 99.47\%; similarly, the Commercial model shows reductions of 98.78\% and 96.51\%, respectively. In contrast, transfer is much weaker for MalConv, where Gamma-padding hardening reduces Gamma-shift ASR by only 0.08\%, while Gamma-shift hardening reduces Gamma-padding ASR by 10.53\%. In contrast, transfer between MAB-Malware and AIMED is substantially less consistent despite their overlapping transformation spaces (Table~\ref{tab:transformations}), as discussed in RQ1. This difference is also reflected in how the two attacks use those transformations: successful MAB-Malware evasions are concentrated on a small subset of actions, while AIMED distributes successful sequences across a much broader portion of its action space. Shared transformation availability alone is insufficient to predict defense transferability; the way transformations are selected and composed during attack generation leads to the results. More generally, Figure~\ref{fig:ari_heatmaps} shows that adversarial hardening can also increase susceptibility to unseen attacks; for example, AIMED-hardened EMBER-17 increases ASR under Gamma-padding and Gamma-shift by 69.90\% and 71.21\%, respectively.

\textbf{Insights from RQ4:}
Our findings suggest that adversarial hardening does not benefit all malware detection paradigms equally. Engineered feature-based detectors exhibit substantially stronger robustness gains than the end-to-end detector considered in our evaluation, indicating that the effectiveness of adversarial retraining depends strongly on the underlying feature representation and model architecture. We further observe that similarity between attacks is not, by itself, a reliable predictor of defense transferability. Although strong reciprocal transfer occurs between Gamma-padding and Gamma-shift for some detectors, the same effect is weak or asymmetric for MalConv, while attacks sharing overlapping functionality-preserving transformations, such as MAB-malware and AIMED, exhibit considerably less consistent transfer. These results motivate further investigation into which properties of adversarial examples determine whether robustness learned against one attack generalizes to others. In practice, adversarial hardening may therefore need to be complemented by defenses that are less dependent on the particular attack distribution observed during training.

%% file: 08_discussion.tex
\section{Limitations and Future Work}
\label{sec:limitations}

Our findings across all research questions reveal that robustness in malware detection is not an inherent property of a model, but 
is shaped by architectural biases, data representation, and the diversity of adversarial manipulations. While this study provides a comprehensive analysis, some limitations must be acknowledged.

First, our work inherits the inherent constraints of problem-space attacks in Windows malware, including computational complexity and data availability. Acquiring original malware and benign executables is challenging due to copyright and safety concerns; consequently, our models were trained on the EMBER dataset, which represents only a subset of the full distribution. 
Second, while we evaluated nine representative problem-space evasion attacks, other published strategies were not reproducible, meaning our analysis may not capture the entire threat landscape of problem-space vulnerabilities. In particular, the public implementation of~\cite{lucas2023adversarial, lucas2021malware, kreuk2018adversarial} is either incomplete or relies on proprietary disassembly tools (e.g., IDA Pro), which challenges their reproducibility.

Taken together, these limitations and our core findings advocate for a paradigm shift: from building attack-specific defenses toward fostering representation-level resilience. Our results indicate that many vulnerabilities stem not from the learning algorithm itself, but from feature representations that allow superficial, semantics-preserving modifications to dictate a model's decision. Therefore, we propose that future research should prioritize the exploration of robust feature representations for Windows malware detection, which are inherently less susceptible to such adversarial manipulations.

%% file: 09_conclusion.tex
\section{Conclusion} 
\label{sec:conclusion}
In this work, we systematically evaluated the vulnerability of Windows malware classifiers to realistic evasion attacks, examining attack effectiveness, cross-model transferability, and adversarial hardening. Our findings reveal significant variability in attack effectiveness and transferability across models and datasets, highlighting important limitations in the robustness of current malware detectors. We also show that evaluating attacks in isolation can underestimate real-world risk, and that adversarial hardening, while beneficial in some cases, does not consistently protect against diverse attack strategies. Overall, our study provides a reproducible benchmark and systematic evaluation methodology that can support the development of more robust malware classifier and more realistic evaluation practices. 

\clearpage

\section*{Open Science}

To support reproducibility, we release the code, experimental configurations, and trained models required to reproduce the reported results. Raw malware binaries will not be redistributed due to safety and licensing considerations.

\section*{LLM Usage Considerations}

LLMs were used for editorial assistance and support with code development and debugging. All generated content was reviewed and validated by the authors. LLMs were not used to generate experimental data or results, and the authors remain fully responsible for the accuracy and integrity of the manuscript.

\section*{Ethical Considerations}
This study analyzes Windows malware detection systems in fully isolated environments, using authorized or public samples with no exposure to live systems. All malware processing and execution is conducted within these isolated research environments, which are separated from production systems and configured to prevent interaction with third-party hosts. We do not deploy malware outside these environments, evaluate attacks against operational security products without authorization, or distribute live malware, generated adversarial binaries, or transformation payloads. The problem-space attacks used in our evaluation are existing techniques from prior work; we do not introduce new functionality-preserving binary transformations. Vulnerabilities discovered during this study were responsibly disclosed to the affected vendors, with the goal of improving adversarial robustness and strengthening malware detection systems.

%% file: Appendix.tex
\clearpage

\appendix
\section{Appendix A}

\subsection{Overview of Evaluated Models and Attack Methods in Prior Work}
\label{subsec:prior_work} 
We provide a structured overview of existing literature on evasion attacks in malware detection, highlighting the attack settings and models evaluated in each study in Table \ref{tab:papers_models}.

\begin{table*}[!t]
\centering
\caption{Overview of adversarial malware attacks and the models or platforms evaluated in prior work, including both academic and commercial detectors}
\label{tab:papers_models}

\resizebox{\textwidth}{!}{%
\small 

\begin{tabular}{lp{6cm}cccccccc}

\toprule
Paper & Attacks & CNN/DNN/MLP & MalConv & Random Forest & LGBM & VirusTotal & AvastNet & FireEye & Commercial \\ 
\midrule
\addlinespace[1pt]  

\cite{tran2022robustness} & Salting & \checkmark &  &  &  &  &  &  & \\

\addlinespace[1pt]  
\cmidrule(lr){2-10}

\cite{yuan2020black} & GAPGAN, Opt., AdvSeq, MalGAN &  & \checkmark &  &  &  &  &  & \\ 

\addlinespace[1pt]  
\cmidrule(lr){2-10}

\cite{zhong2023malfox} & MalFox: Obfusmal, Stealmal, Hollowmal &  &  &  &  & \checkmark &  &  & \\ 

\addlinespace[1pt]  
\cmidrule(lr){2-10}

\cite{do2022method} & RL-based method &  & \checkmark & \checkmark & \checkmark &  &  &  & \\ 

\addlinespace[1pt]  
\cmidrule(lr){2-10}

\cite{chen2019adversarial} & FGSM, Random, and Experience-based method &  & \checkmark &  &  &  &  &  & \\ 

\addlinespace[1pt]  
\cmidrule(lr){2-10}

\cite{lucas2021malware} & Binary diversification &  & \checkmark &  &  &  & \checkmark &  & \\

\addlinespace[1pt]  
\cmidrule(lr){2-10}

\cite{anderson2017evading} & gym-malware &  &  &  & \checkmark &  &  &  &  \\

\addlinespace[1pt]  
\cmidrule(lr){2-10}

\cite{wu2018enhancing} & Gym-plus, gym-malware &  & \checkmark &  & \checkmark &  &  & \checkmark &  \\

\addlinespace[1pt]  
\cmidrule(lr){2-10}

\cite{ebrahimi2021binary} & RL-based (BFA, DDQN, ACER, AMG-VAC) &  & \checkmark &  & \checkmark &  &  & \checkmark &  \\

\addlinespace[1pt]  
\cmidrule(lr){2-10}

\cite{labaca-castro2021aimed-rl} & AIMED-RL (DDQN, DuDDQN, ACER, DiDDQN) &  &  &  & \checkmark &  &  &  &  \\ 

\addlinespace[1pt]  
\cmidrule(lr){2-10}

\cite{labaca-castro2019armed} & ARMED &  &  &  &  &  &  &  & \checkmark  \\

\addlinespace[1pt]  
\cmidrule(lr){2-10}

\cite{song2020automatic} & RL-based &  &  &  & \checkmark &  &  &  & \checkmark \\

\addlinespace[1pt]  
\cmidrule(lr){2-10}

\cite{labaca-castro2019aimed} & AIMED, ARMED &  &  &  & \checkmark &  &  &  & \checkmark \\

\addlinespace[1pt]  
\cmidrule(lr){2-10}

\cite{ceschin2019shallow} & Append strings from goodware \& Packing  &  & \checkmark  &  & \checkmark &  &  &  &  \\ 

\addlinespace[1pt]  
\cmidrule(lr){2-10} 

\cite{wang2020mdea} & MDEA &  & \checkmark &  &  &  &  &  &  \\ 

\addlinespace[1pt]  
\cmidrule(lr){2-10}

\cite{demetrio2021functionality} & GAMMA attacks &  & \checkmark &  & \checkmark &  &  &  &  \\

\addlinespace[1pt]  
\cmidrule(lr){2-10}

\cite{fleshman2018static} & binary-search based &  & \checkmark &  &  &  &   &  & \checkmark \\

\addlinespace[1pt]  
\cmidrule(lr){2-10}

\cite{suciu2019exploring} & FGM Append, Slack FGM &  & \checkmark &  &  &  &  &  &  \\

\addlinespace[1pt]  
\cmidrule(lr){2-10} 

\cite{kozak2024creating} & AMG-PPO, AMG-random, MAB-Malware &  & \checkmark \footnote{only evaluated for transferability} &  & \checkmark &  &  &  &  \\

\addlinespace[1pt]  
\cmidrule(lr){2-10}

\cite{imran2024evaluating} & Extend, Full DOS, Shift, FGSM padding + slack, GAMMA &  & \checkmark &  & \checkmark &  &  &  &  \\

\addlinespace[1pt]  
\cmidrule(lr){2-10}

\cite{louthanova2024comparison} & Partial DOS, Full DOS, GAMMA, Gym-malware &  & \checkmark &  & \checkmark &  &  &  & \checkmark \\

\cmidrule(lr){2-10}

\cite{fang2019evading} & DQEAF & &  &  & \checkmark &  &  &  &  \\

\cmidrule(lr){2-10}

\cite{zhong2022reinforcement} & MalInfo, MalFox & &  &  &  & \checkmark &  &  &  \\

\cmidrule(lr){2-10}

\cite{wang2022black} & AMG-PDG & &  \checkmark &  & \checkmark &  &  &  &  \\



\cmidrule(lr){2-10}
 \cite{zhan2023malpatch} & MalPatch, Random Patch,
Benign Patch and Transfer Patch, GAME-UP and BASGAN  & \checkmark & \checkmark & & & & \checkmark & \\

\cmidrule(lr){2-10}

\cite{li2025minimal} & MiniMal, GAMMA and MAB-malware & & \checkmark  \checkmark \footnote{evaluates both versions of MalConv} & & \checkmark & & \checkmark & & \checkmark \\ 

\cmidrule(lr){2-10}

\cite{sang2025generating} & GanGenetic, MalGAN, Mab-malware, and GAMMA & \checkmark & \checkmark & \checkmark & \checkmark & & & &  \\

\cmidrule(lr){2-10}
 
\cite{de2025olivander} & OLIVANDER, AMG and GAMMA & \checkmark & & & &  & & &  \\ 

\cmidrule(lr){2-10}

\cite{anderson2018learning} & RL-based &  & & & \checkmark & & & &  \\

\cmidrule(lr){2-10}

\cite{song2020mab} & MAB-Malware, SecML-Malware and Gym-Malware &  & \checkmark & &  \checkmark & & & &  \checkmark \\

\cmidrule(lr){2-10}

\cite{fang2020deepdetectnet} & RLAttackNet & \checkmark & & & & & & & \\

\cmidrule(lr){2-10}

\cite{quertier2022merlin} & REINFORCE, DQN  &  & \checkmark & &  \checkmark & & & & \checkmark \\

\cmidrule(lr){2-10}

\cite{ebrahimi2020binary} & MalRNN  & \checkmark & \checkmark \checkmark  & &  & & & & \\

\bottomrule

\end{tabular}
}
\end{table*}

\subsection{Problem‑Space binary transformations and obfuscation techniques}
\label{subsec:techniques} 

We define the problem‑space transformations and obfuscation techniques used by the evaluated attacks. All transformations are applied in the problem space and, when used by an attack, are constrained to preserve runtime executability and functionality (see Section~\ref{subsec:sandboxing}).

\begin{enumerate}[leftmargin=0pt]
\item \textbf{Gamma‑Padding.} Appends benign or innocuous bytes (an overlay) to the end of the PE file so that the file size and byte‑level distribution change while the original code and headers remain intact.
\item \textbf{Gamma‑Shift.} Shifts section contents and inserts benign padding inside or between sections, adjusting offsets so that payload bytes are relocated rather than simply appended to the file end.
\item \textbf{Gamma‑Sections.} Adds one or more new named PE sections containing benign data and updates the section table and headers accordingly, thereby modifying the binary’s structural metadata (section names, offsets, and sizes).
\item \textbf{MiniMal.} The MiniMal attack uses 4 transformations from Table \ref{tab:transformations} including  MD, AS, SA, and padding. 
\item \textbf{MAB-malware.} The MAB-malware attack uses 7 transformations from Table \ref{tab:transformations} including  OA, SR, SA, SP, RC, RD and BC. 
\item \textbf{Other attacks.} The remaining attacks select from a predefined set of 10 problem‑space transformations OA, IP, SR, SA, SP, RS, RD, UP, UPD, and BC from (Table~\ref{tab:transformations}) according to each attack's search strategy and constraints.
\end{enumerate}

\newcolumntype{L}{>{\RaggedRight\arraybackslash}X} 


\begin{table*}[!htbp]
\centering
\caption{Problem-space binary transformations and obfuscation techniques.}
\label{tab:transformations}
\begin{tabularx}{\linewidth}{@{}lL@{}}
\toprule
\textbf{Technique - Abbreviation} &  \textbf{Description} \\
\midrule
Overlay Append - OA & Appends benign contents to the end of a binary (overlay). Typically preserves functionality and alters byte-based features. \\
Section Append - SP & Writes bytes into unused space within an existing PE section; low risk but can be fragile if offsets are miscomputed. \\
Section Add - SA & Adds a new named PE section containing benign content; modifies PE headers and section tables and can break binaries if not done correctly. \\
Section Rename - SR & Renames sections to common benign names to evade simple heuristics. \\
Remove Certificate - RC & Zeroes out or removes the certificate block (affects signing metadata). \\
Remove Debug - RD & Strips debug information (reduces static cues about build environment). \\
Remove Signature - RS & Strips or invalidates code signatures (alters provenance signals). \\
Imports Append - IP & Adds unused or benign API imports to the import table to change import-based features. \\
Break Checksum - BC &Zeroes the optional header checksum, altering PE integrity flags (may affect some loaders). \\
Code Randomization - CR &  Rewrites instruction sequences to semantically equivalent variants (true code obfuscation). \\
UPX Compression/Packing - UP & Packs the binary with UPX (packing is a form of obfuscation; requires an unpacking stub at runtime). \\
UPX Decompression/Unpacking - UPD & Removes UPX packing (not an obfuscation, but useful for analysis/unpacking). \\
Modify DOS Header - MD & Modifies or injects bytes in the DOS header padding between the “MZ” signature and the PE header without affecting execution. \\
Slack Append - AS &  Appends bytes into section slack space (alignment padding between PE sections) without affecting program execution. \\

\bottomrule
\end{tabularx}
\end{table*}

\subsection{Datasets}
\label{subsec:datasets_appendix}

Table~\ref{tab:datasets_summary} summarizes the datasets used in our experiments, reporting the feature type, collection period, dataset size, train-test splits, and data availability, thereby highlighting differences in scale, temporal coverage, and representation. Additionally, Table~\ref{tab:dataset} provides, for each dataset and split, the number of executable samples we were able to retrieve for our experiments alongside the original published counts (``collected / original'').

{

\renewcommand{\arraystretch}{0.5} 

\newcolumntype{L}{>{\RaggedRight\arraybackslash}X} 
\begin{table*}[t]
\footnotesize
\centering
\caption{Summary of datasets used in our experiments.}
\label{tab:datasets_summary}
\begin{tabularx}{\textwidth}{@{}L L L L L L@{}}

\toprule
\textbf{Dataset} & \textbf{Features} & \textbf{Collected Year} & \textbf{Size} & \textbf{Train-Test Split} & \textbf{Availability} \\
\midrule
SOREL-20m & EMBER-v2 features & Jan 2017 – Apr 2019 & 9,919,251 malware, 9,470,626 benign & Training: 65.49\%, Validation: 12.87\%, Test: 21.64\% & Extracted features and disarmed malware binaries\\

EMBER 2017 & EMBER-v1 features & in or before 2017 & 400K malware, 400K benign, 300K unlabeled & Train: 81.82\% and Test: 18.18\% & Extracted features \\

EMBER 2018 & EMBER-v2 features & in or before 2018 & 400K malware, 400K benign, 200K unlabeled & Train: 80\% and Test: 20\% & Extracted features \\

EMBER 2024 & EMBER-v3 features & Sep, 2023 – Sep, 2024 & 1,616,000 malware, 1,616,000 benign, and 6,291 challenge & Training: ~81.3\%  and Test: ~18.8\%  & Extracted features \\

\bottomrule
\end{tabularx}
\end{table*}
}

Acquiring executable samples is a non-trivial step for problem-space evaluations because several public datasets only publish extracted features, images, or disarmed binaries rather than runnable PE files. Hence to create the NTS dataset and to obtain runnable binaries we used a hash-based search strategy: for each sample in the feature-only datasets (e.g., EMBER v1, v2, and v3 and SOREL-20M) we searched public malware repositories and archival services (e.g., Hybrid Analysis \cite{hybrid}, VirusShare \cite{virusshare.com_2020}, \cite{aghakhani2020malware}), and downloaded matching binaries when available. For example, EMBER-2017 originally provides 300k malware and 300k benign samples in the training split; from this set we recovered 51,352 malware and 43,983 benign binaries for our training experiments. Similarly, for EMBER-2018 we recovered 2,500 malware and 2,500 benign samples from the original test split of 100k / 100k. In contrast, the proprietary Commercial dataset was available to us as full binaries. SOREL-20M presents a particular challenge: the public release includes a large collection of disarmed (non-executable) malware samples and extracted features; disarmed binaries are unsuitable for executability-preserving problem-space attacks because they cannot be executed in a sandbox environment. 
Therefore, for SOREL we again used the hash-based search strategy to recover runnable binaries and constructed a balanced test subset (reported in Table~\ref{tab:dataset}) that preserves the temporal and class distributions recommended by the dataset authors \cite{harang2020sorel}.


\begin{table}[ht!]
\centering
\caption{Collected vs Original Samples across datasets.}
\label{tab:dataset} 
\begin{tabular}{llccc} 
\hline
Dataset & Split & Malware & Benign \\
\hline
EMBER 2017 & Train & 51,352 / 300,000 & 43,983 / 300,000 \\
EMBER 2017 & Test  & 12,500 / 100,000 & 12,500 / 100,000  \\
EMBER 2018 & Test  & 2,500 / 100,000  & 2,500 / 100,000  \\
SOREL-20m & Test  & 12,501   / 1,360,622 &   22,282  / 2,834,441 \\
Commercial & Train & 60,397 / 60,397 & 53,084 / 53,084  \\
Commercial & Test  & 1,670/ 1,670 & 1,055 / 1,055 \\ 
EMBER 2024 & Test &  17,345 / 270,000 & 12,441 / 270,000 \\
\hline
\end{tabular}
\end{table}

\subsection{Model performance on clean samples}
\label{clean_performance_of_models}

We report threshold-dependent performance metrics for all evaluated malware detectors on clean (non-adversarial) samples in \ref{tab:model_clean_performance}, including false positive rate (FPR), recall (true positive rate), precision, F1-score, accuracy, and AUC. Following common practice in the malware detection literature, we first calibrate all models at a fixed FPR of 0.1\%. However, for models where this operating point results in a recall (TPR) below 70\%, we relax the constraint and report results at an FPR of 1.0\% to ensure a minimally functional detection regime. Overall, most models achieve high AUC values (>0.95), indicating strong ranking ability, while performance differences are primarily reflected in recall–precision trade-offs induced by threshold selection. In particular, EMBER-based models and MalConv variants achieve the highest F1-scores and recall at low FPR settings, whereas the commercial model exhibits comparatively lower accuracy despite strong precision, suggesting a more conservative detection strategy.

\begin{table}[h!]
\centering
\caption{Comparison of clean malware detectors with respect to threshold-dependent metrics: recall, precision, F1-score, accuracy, and AUC.}
\label{tab:model_clean_performance}
\resizebox{\columnwidth}{!}{

\begin{tabular}{lccccccc}
\toprule
\textbf{Model} & \textbf{Threshold} & \textbf{FPR} & \textbf{Recall (TPR)} & \textbf{Precision} & \textbf{F1-score} & \textbf{Accuracy} & \textbf{AUC} \\
\midrule
MalConv & 0.9940    & 0.1\% & 0.8166 & 0.9988 & 0.8985 & 0.9078 & 0.9959 \\
MalConv2 & 0.997622 & 1.0\% & 0.70 & 0.9868 & 0.8190 & 0.9322 & 0.9775 \\
EMBER-17  & 0.8710    & 0.1\% & 0.9300 & 0.9989 & 0.9632 & 0.9645 & 0.9991 \\
EMBER-18  & 0.9996   & 0.1\% & 0.8681 & 0.9989 & 0.9289 &  0.9335 & 0.9964 \\
EMBER-24 & 0.9557   & 0.1\% & 0.8850 & 0.9989 & 0.9385 & 0.9420 & 0.9982 \\ 
FFNN-SOREL & 0.9352 & 0.1\% & 0.8450 & 0.9867 & 0.9100 & 0.9397 & 0.9884 \\
LGBM-SOREL & 0.9492 & 1.0\% & 0.7457 & 0.9678 & 0.8414 & 0.9006 & 0.9491 \\
Commercial model & 0.9741   & 1.0\% & 0.7214 & 0.9918 & 0.8352 & 0.8256 & 0.9831 \\ 
\bottomrule
\end{tabular}
}
\end{table}

\subsection{Attack results (RQ1)} 
\label{subsec:attack_results} 
Table~\ref{tab:asr_with_func} compares ASR computed on the raw set of generated evasive samples (``w/o exec.'') with ASR computed after we filter out non-executable samples using our sandbox-based checks (``w. exec.''). For each model–attack pair the ``w/o exec.'' row reports the ASR obtained by counting every generated sample that evaded the detector, whereas the ``w. exec.'' row reports the ASR after retaining only those evasive samples that also passed our executability tests. 

Two points are worth emphasizing. First, enforcing executability substantially reduces the measured effectiveness of some attacks. For example, the Gamma-section attack suffers a dramatic drop in ASR once non-executable samples are removed, indicating that many of its reported evasions produce non-operational binaries. In contrast, the other Gamma variants (shift and padding) exhibit only minor differences between ASR before and after executability checks.

The reason is structural: Gamma-section modifies the PE file’s section table and other header fields, inserting new named sections and altering offsets that determine in-memory layouts, imports, and resources. Even small inconsistencies can cause parsers or the Windows loader to fail. By comparison, Gamma-padding appends data at the end of the file without affecting loader-visible metadata, and Gamma-shift makes minor in-file relocations that typically preserve functionality and executability. This structural fragility explains why many Gamma-section binaries fail executability checks, reducing the measured ASR.

\begin{table*}[ht!]
\centering
\caption{Mean ASR (\%) with and without executability evaluation across models. \textbf{Bold} and \underline{underlined} values indicate the highest and second-highest ASR, respectively, considering only executability-preserving adversarial examples.}
\label{tab:asr_with_func}
\resizebox{\textwidth}{!}{
\small
\begin{tabular}{lcccccccccc}
\toprule
\textbf{Model} 
& \multicolumn{2}{c}{\textbf{MAB-malware}} 
& \multicolumn{2}{c}{\textbf{Gamma-shift}} 
& \multicolumn{2}{c}{\textbf{Gamma-padding}} 
& \multicolumn{2}{c}{\textbf{Gamma-section}} 
& \multicolumn{2}{c}{\textbf{MiniMal}} \\  
\cmidrule(lr){2-3} 
\cmidrule(lr){4-5} 
\cmidrule(lr){6-7} 
\cmidrule(lr){8-9} 
\cmidrule(lr){10-11}

& w/o exec. & w. exec. 
& w/o exec. & w. exec. 
& w/o exec. & w. exec. 
& w/o exec. & w. exec. 
& w/o exec. & w. exec. \\

\midrule

FFNN-SOREL 
& 85.63 
& \textbf{85.63}
& 15.97 
& 15.97 
& 14.17 
& 14.17 
& 41.17 
& 3.27 
& 65.20 
& \underline{65.20} \\

LGBM-SOREL 
& 95.30 
& \textbf{95.30} 
& 70.50 
& 70.50 
& 70.40 
& 70.40 
& 26.93 
& 3.93 
& 81.50 
& \underline{81.50} \\

EMBER-17 
& 89.03 
& \textbf{89.03} 
& 48.83 
& 48.83 
& 48.87 
& 48.87 
& 72.23 
& 5.43 
& 64.80 
& \underline{64.80} \\

EMBER-18 
& 98.43 
& \textbf{98.33} 
& 91.93 
& \underline{91.93} 
& 91.10 
& 91.00 
& 96.70 
& 4.60 
& 77.40 
& 77.40 \\

EMBER-24 
& 83.40 
& \textbf{83.40} 
& 12.73 
& 12.73 
& 7.33 
& 7.33 
& 0 
& 0 
& 30.60 
& \underline{30.60} \\

MalConv 
& 98.93 
& \underline{98.83} 
& 100.00 
& \textbf{100.00} 
& 98.63 
& 98.53 
& 99.83 
& 4.53 
& 98.30 
& 98.20 \\

MalConv2 
& 98.80 
& \textbf{98.70} 
& 88.53 
& \underline{88.53} 
& 82.07 
& 81.97 
& 90.60 
& 2.20 
& 11.00  
& 11.00 \\

Commercial 
& 100.00 
& \textbf{99.90} 
& 97.80 
& \underline{97.80} 
& 91.23 
& 91.13 
& 92.70 
& 3.20 
& 97.36 
& 97.23 \\

\cmidrule(lr){1-11}

Avg  
& 93.69 
& \textbf{93.64} 
& 65.79 
& \underline{65.79} 
& 62.98 
& 62.93 
& 65.02 
& 3.40 
& 65.77 
& 65.74 \\

\bottomrule
\end{tabular}
}
\end{table*}

\subsection{Attack Effectiveness on the NTS dataset (RQ1)}
\label{subsec:attack_effectiveness}

\begin{table*}[ht!]
\centering
\caption{ASR\% across models on the NTS dataset. For Gamma and MAB attacks, results are shown without (w/o) and with (w.) executability evaluation. \textbf{Bold} indicates highest ASR per row; \underline{underline} indicates second-highest.}
\label{tab:combined_asr}
\resizebox{\textwidth}{!}{
\small
\begin{tabular}{lccccccccccccccc}
\toprule
\textbf{Model} 
& \multicolumn{2}{c}{\makecell{\textbf{MAB-}\\\textbf{malware}}} 
& \multicolumn{2}{c}{\makecell{\textbf{Gamma-}\\\textbf{shift}}} 
& \multicolumn{2}{c}{\makecell{\textbf{Gamma-}\\\textbf{padding}}} 
& \multicolumn{2}{c}{\makecell{\textbf{Gamma-}\\\textbf{section}}} 
& \multicolumn{2}{c}{\makecell{\textbf{MiniMal}}}
& \makecell{\textbf{AIMED}} 
& \makecell{\textbf{AIMED-}\\\textbf{RL}} 
& \makecell{\textbf{ARMED}} 
& \makecell{\textbf{Game-}\\\textbf{up}} \\

\cmidrule(lr){2-3} \cmidrule(lr){4-5} \cmidrule(lr){6-7} \cmidrule(lr){8-9}  \cmidrule(lr){10-11}

& w/o & w. 
& w/o & w. 
& w/o & w. 
& w/o & w. 
& w/o & w. 
&  &  &  &  \\

\midrule

FFNN-SOREL 
& 57.3 & \textbf{55.3}
& 15.1 & 14.8
& 14.2 & 13.9 
& 36.0 & 3.5 
& 20.3 & \underline{18.0}
& 6.3 & 4.0 & 0.8 & 1.4 \\

LGBM-SOREL 
& 99.0 & \textbf{87.7} 
& 34.7 & 33.5
& 35.0 & 33.4 
& 25.4 & 3.0 
& 54.4 & \underline{52.1}
& 14.2 & 12.8 & 4.9 & 9.1 \\

EMBER-17 
& 84.8 & \textbf{84.5} 
& 54.0 & 53.7
& 52.2 & 51.9 
& 68.6 & 21.8 
& 70.5 & \underline{70.4} 
& 27.2 & 21.9 & 20.9 & 8.7 \\

EMBER-18 
& 99.8 & \textbf{99.7}
& 97.9 & 97.8
& 97.9 & 97.8
& 97.9 & 3.7 
& 98.0 & \underline{98.0} 
& 30.9 & 33.7 & 8.8 & 5.1 \\

EMBER-24 
& 9.6 & \textbf{9.6}
& 8.1 & \underline{8.1}
& 5.6 & 5.6 
& 0 & 0 
& 4.6 & 4.5 
& 1.4 & 0.4 & 0.2 & 0 \\

MalConv 
& 65.2 & 65.1 
& 100.00 & \textbf{99.6} 
& 60.3 & 60.2 
& 94.7 & 21.6 
& 77.6 & \underline{77.4}
& 29.2 & 12.8 & 6.5 & 23.3 \\

MalConv2 
& 94.9 & \textbf{94.8} 
& 82.3 & \underline{82.1} 
& 78.2 & 78.0 
& 92.1 & 4.1 
& 12.3 & 12.3 
& 58.9 & 47.6 & 15.5 & 51.2 \\

Commercial 
& 99.7 & \textbf{99.1} 
& 94.6 & 94.0 
& 93.6 & 93.0 
& 94.6 & 28.6 
& 98.5 & \underline{97.6}
& 33.2 & 25.1 & 4.6 & 19.6 \\

\midrule

Avg 
& 76.29 & \textbf{74.48}
& 60.84 & 60.45
& 54.63 & 54.23
& 63.66 & 10.79
& 54.53 & \underline{53.79}
& 25.16 & 19.79 & 7.78 & 14.80 \\

\bottomrule
\end{tabular}
}
\end{table*}

\begin{table*}[ht!]
\centering
\caption{Coverage of other attacks by the samples evaded by MAB-malware across models. Each attack column is divided into two sub-columns: \texttt{Overlapping / Evasive} and \texttt{Coverage (\%)}.} \label{tab:coverage_matrix_mab_reordered}

\resizebox{\textwidth}{!}{%
\small
\begin{tabular}{lll|ll|ll|ll|ll|ll|ll|ll}
\toprule

\textbf{Model} 
& \multicolumn{2}{c|}{\makecell{\textbf{Gamma-}\\\textbf{shift}}} 
& \multicolumn{2}{c|}{\makecell{\textbf{Gamma-}\\\textbf{padding}}} 
& \multicolumn{2}{c|}{\makecell{\textbf{Gamma-}\\\textbf{section}}}  
& \multicolumn{2}{c|}{\textbf{MiniMal}}
& \multicolumn{2}{c|}{\textbf{AIMED}}
& \multicolumn{2}{c|}{\textbf{AIMED-RL}}
& \multicolumn{2}{c|}{\textbf{ARMED}}
& \multicolumn{2}{c}{\makecell{\textbf{Game-}\\\textbf{Up}}} \\

\midrule

FFNN-SOREL 
& 146 / 160 & 91.25\%
& 142 / 142 & 100\%
& 26 / 30 & 86.67\%
& 646 / 652 & 99.08\%
& 32 / 50 & 64\%
& 5 / 5 & 100\%
& 6 / 6 & 100\%
& 0 / 0 & \textemdash \\

LGBM-SOREL
& 692 / 705 & 98.16\%
& 700 / 704 & 99.43\%
& 40 / 40 & 100\%
& 805 / 815 & 98.77\%
& 162 / 162 & 100\%
& 68 / 76 & 89.47\%
& 48 / 52 & 92.31\%
& 33 / 44 & 75\% \\

EMBER-2017
& 476 / 488 & 97.54\%
& 481 / 488 & 98.57\%
& 43 / 54 & 79.63\%
& 642 / 648 & 99.07\%
& 8 / 9 & 88.89\%
& 2 / 2 & 100\%
& 1 / 2 & 50\%
& 3 / 3 & 100\% \\

EMBER-2018
& 919 / 919 & 100\%
& 910 / 910 & 100\%
& 46 / 46 & 100\%
& 774 / 774 & 100\%
& 563 / 563 & 100\%
& 305 / 305 & 100\%
& 195 / 195 & 100\%
& 260 / 260 & 100\% \\

EMBER-24
& 118 / 127 & 92.91\%
& 72 / 73 & 98.63\%
& 0 / 0 & \textemdash
& 304 / 306 & 99.35\%
& 17 / 17 & 100\%
& 3 / 3 & 100\%
& 1 / 4 & 25\%
& 0 / 0 & \textemdash \\

MalConv
& 989 / 1000 & 98.90\%
& 985 / 986 & 99.90\%
& 44 / 46 & 95.65\%
& 980 / 982 & 99.80\%
& 224 / 224 & 100\%
& 140 / 143 & 97.90\%
& 131 / 136 & 96.32\%
& 117 / 125 & 93.60\% \\

MalConv2
& 882 / 885 & 99.66\%
& 819 / 820 & 99.88\%
& 21 / 22 & 95.45\%
& 110 / 110 & 100\%
& 351 / 351 & 100\%
& 220 / 222 & 99.10\%
& 190 / 197 & 96.45\%
& 254 / 258 & 98.45\% \\

Commercial
& 978 / 978 & 100\%
& 911 / 911 & 100\%
& 31 / 32 & 96.88\%
& 973 / 973 & 100\%
& 157 / 157 & 100\%
& 77 / 77 & 100\%
& 83 / 83 & 100\%
& 95 / 95 & 100\% \\

\midrule

\textbf{Total}
& 5200 / 5262 & 98.82\%
& 5020 / 5034 & 99.72\%
& 251 / 270 & 92.96\%
& 5234 / 5260 & 99.51\%
& 1514 / 1533 & 98.76\%
& 820 / 833 & 98.44\%
& 655 / 675 & 97.04\%
& 762 / 785 & 97.07\% \\

\bottomrule
\end{tabular}
}
\end{table*}

\subsection{Coverage of MAB-malware (RQ2)} 
\label{subsec:mab_coverage} 
Table \ref{tab:coverage_matrix_mab_reordered} presents the coverage of other attacks by samples that successfully evade detection through the MAB-malware attack. Although MAB-malware shows high overlap with Gamma-padding (99.57\%), MiniMal (99.34\%), and Gamma-shift (98.44\%), its coverage notably decreases for Gamma-section (93.37\%), Game-Up (96.79\%), and ARMED (96.93.45\%). This suggests that while MAB-malware is highly representative of attacks from the Gamma family, it misses many samples evaded by other attack types. These results highlight that MAB-malware does not fully span the evasive behavior space.

\subsection{Performance of Hardened Models on clean samples (RQ4)}
\label{hardened_model_clean_performance}

In \ref{tab:hardened_accuracy}, we report the performance of hardened models under different defense mechanisms at a fixed false-positive rate (FPR) operating point. Following the same evaluation practice as the original models, most models are calibrated at an FPR of 0.1\%, while for the commercial model we use a relaxed operating point of 1.0\% due to significantly reduced TPR under stricter constraints. We evaluate recall, precision, F1-score, accuracy, and AUC across different defense strategies, including Gamma-padding, Gamma-shift, MAB-malware, and AIMED, to assess their impact on classification performance.

\begin{table*}[ht]
\centering
\caption{Comparison of hardened model performance at a fixed FPR operating point, reporting threshold, classification metrics, and AUC}
\label{tab:hardened_accuracy}

\resizebox{\textwidth}{!}{

\begin{tabular}{llcccccccc}
\toprule
\textbf{Model} & \textbf{Hardened Against} & \textbf{Threshold} & \textbf{FPR} & \textbf{Recall (TPR)} & \textbf{Precision} & \textbf{F1-score} & \textbf{Accuracy} & \textbf{AUC} \\
\midrule
\multirow{5}{*}{\textbf{EMBER-17}} 

& No defense &  0.8710 & 0.1\% & 0.9300 & 0.9989 & 0.9632 & 0.9645 & 0.9991  \\
& Gamma-padding & 0.9451 & 0.1\%  & 0.9879 & 0.9990 & 0.9879 & 0.9934 & 0.9998\\
& Gamma-shift & 0.9443 & 0.1\% & 0.9878 & 0.9990 & 0.9934 & 0.9934 & 0.9997 \\
& MAB-malware & 0.9681 & 0.1\% & 0.9878 & 0.9990 & 0.9934 & 0.9934 & 0.9998 \\
& MiniMal & 0.9559 & 0.1\% & 0.9870 & 0.9990 & 0.9930 & 0.9930 & 0.9997 \\
& AIMED & 0.9494 & 0.1\% & 0.9876 & 0.9990 & 0.9932 & 0.9933 & 0.9998 \\

    \cmidrule(lr){2-9}

\multirow{5}{*}{\textbf{MalConv}} 

& No defense & 0.9940 & 0.1\% & 0.8166 & 0.9988 & 0.8985 & 0.9078 & 0.9959 \\
& Gamma-padding & 0.9883 & 0.1\% & 0.8209 & 0.9988 & 0.9012 & 0.9100 & 0.9951 \\
& Gamma-shift & 0.9931 & 0.1\% & 0.8011 & 0.9989 & 0.8891 & 0.9001 & 0.9953 \\
& MAB-malware & 0.9874 & 0.1\% & 0.8500 & 0.9989 & 0.9184 & 0.9245 & 0.9951 \\
& MiniMal & 0.9921 & 0.1\% & 0.8258 & 0.9988 & 0.9041 & 0.9124 & 0.9957 \\
& AIMED & 0.9912 & 0.1\% & 0.8158 & 0.9988 & 0.8981 & 0.9074 & 0.9954 \\
    \cmidrule(lr){2-9}

\multirow{5}{*}{\textbf{Commercial}}

& No defense & 0.9741 & 1.0\% & 0.7214 & 0.9918 & 0.8352 & 0.8256 & 0.9831
 \\
& Gamma-padding & 0.9636 & 1.0\% & 0.7310 & 0.9919 & 0.8417 & 0.8315 & 0.9826 \\
& Gamma-shift & 0.9615 & 1.0\% & 0.7244 & 0.9918 & 0.8373 & 0.8275 & 0.9812 \\
& MAB-malware & 0.9840 & 1.0\% & 0.6339 & 0.9906 & 0.7731 & 0.7720 & 0.9801 \\
& MiniMal & 0.9848 & 1.0\% & 0.6201 & 0.9904 & 0.7627 & 0.7636 & 0.9792 \\
& AIMED & 0.9771 & 1.0\% & 0.7232 & 0.9918 & 0.8363 & 0.8267 & 0.9834 \\

\bottomrule
\end{tabular}
}
\end{table*}

\subsection{Performance of Hardened Model on the NTS dataset (RQ4)}
\label{subsec:hardened_model_performance} 

\begin{table*}[ht!] 
\centering 
\scriptsize 
\setlength{\tabcolsep}{2pt} 
\caption{${\Delta \mathrm{ASR}_{\mathrm{abs}}}$ and ${\Delta \mathrm{ASR}_{\mathrm{rel}}}$ before and after adversarial hardening on the NTS dataset, illustrating differences in robustness and defense transfer across attacks.} 
\label{tab:robustness} 
\resizebox{\columnwidth}{!}{

\begin{tabular}{lllcccc} 
\toprule 
\textbf{Model} & 
\textbf{Hardened Against} & 
\textbf{Test Attack} &
$\boldsymbol{\mathrm{ASR}_{\mathrm{before}}}$ & 
$\boldsymbol{\mathrm{ASR}_{\mathrm{after}}}$ &
$\boldsymbol{\Delta \mathrm{ASR}_{\mathrm{abs}}}$ & 
$\boldsymbol{\Delta \mathrm{ASR}_{\mathrm{rel}}}$ (\%) \\ 
\midrule

\multirow{25}{*}{\textbf{EMBER-17}} 
& \multirow{5}{*}{Gamma-padding} 

& \textbf{Gamma-padding} & \textbf{51.9} & \textbf{0.2} 
& \textbf{\good{-51.70}} & \textbf{\good{-99.61\%}} \\ 

& & Gamma-shift & 53.7 & 0.3 
& \good{-53.40} & \good{-99.44\%} \\ 

& & MAB-malware & 84.5 & 41.8 
& \good{-42.70} & \good{-50.53\%} \\

& & MiniMal & 70.4 & 47.0 
& \good{-23.40} & \good{-33.24\%} \\

& & AIMED & 27.2 & 20.1 
& \good{-7.10} & \good{-26.10\%} \\ 

\cmidrule(lr){2-7} 

& \multirow{5}{*}{Gamma-shift} 

& Gamma-padding & 51.9 & 0.6 
& \good{-51.30} & \good{-98.84\%} \\ 

& & \textbf{Gamma-shift} & \textbf{53.7} & \textbf{0.2} 
& \textbf{\good{-53.50}} & \textbf{\good{-99.63\%}} \\ 

& & MAB-malware & 84.5 & 49.5 
& \good{-35.00} & \good{-41.42\%} \\ 

& & MiniMal & 70.4 & 39.7 
& \good{-30.70} & \good{-43.61\%} \\

& & AIMED & 27.2 & 20.0 
& \good{-7.20} & \good{-26.47\%} \\ 

\cmidrule(lr){2-7} 

& \multirow{5}{*}{MAB-malware} 

& Gamma-padding & 51.9 & 0.6 
& \good{-51.30} & \good{-98.84\%} \\ 

& & Gamma-shift & 53.7 & 1.8 
& \good{-51.90} & \good{-96.65\%} \\ 

& & \textbf{MAB-malware} & \textbf{84.5} & \textbf{26.2} 
& \textbf{\good{-58.30}} & \textbf{\good{-68.99\%}}  \\ 

& & MiniMal & 70.4 & 10.8 
& \good{-59.60} & \good{-84.66\%} \\

& & AIMED & 27.2 & 9.8 
& \good{-17.40} & \good{-63.97\%} \\ 

\cmidrule(lr){2-7} 

& \multirow{5}{*}{MiniMal} 

& Gamma-padding & 51.9 & 1.40 
& \good{-50.50} & \good{-97.30\%} \\ 

& & Gamma-shift & 53.7 & 1.8
& \good{-51.9} & \good{-96.64\%} \\ 

& & MAB-malware & 84.5 & 64.0 
& \good{-20.50} & \good{-24.26\%} \\ 

& & \textbf{MiniMal} & \textbf{70.4} & \textbf{0} 
& \textbf{\good{-70.40}} & \textbf{\good{-100.00\%}} \\ 

& & AIMED & 27.2 & 7.0
& \good{20.2} & \good{74.26\%} \\ 

\cmidrule(lr){2-7} 

& \multirow{5}{*}{AIMED} 

& Gamma-padding & 51.9 & 51.2 
& \good{-0.70} & \good{-1.35\%} \\ 

& & Gamma-shift & 53.7 & 53.1 
& \good{-0.60} & \good{-1.12\%} \\ 

& & MAB-malware & 84.5 & 62.5 
& \good{-22.00} & \good{-26.04\%} \\ 

& & MiniMal & 70.4 & 35.7 
& \good{-34.70} & \good{-49.29\%} \\

& & \textbf{AIMED} & \textbf{27.2} & \textbf{0} 
& \textbf{\good{-27.20}} & \textbf{\good{-100.00\%}} \\ 

\cmidrule(lr){1-7}

\multirow{25}{*}{\textbf{MalConv}} 
& \multirow{5}{*}{Gamma-padding} 

& \textbf{Gamma-padding} & \textbf{60.2} & \textbf{43.3} 
& \textbf{\good{-16.90}} & \textbf{\good{-28.07\%}} \\ 

& & Gamma-shift & 99.6 & 99.4 
& \good{-0.20} & \good{-0.20\%} \\ 

& & MAB-malware & 65.1 & 66.6 
& \bad{1.50} & \bad{2.30\%} \\ 

& & MiniMal & 90.8 & 67.9 
& \good{-22.90} & \good{-25.22\%} \\

& & AIMED & 29.2 & 30.4 
& \bad{1.20} & \bad{4.11\%} \\ 

\cmidrule(lr){2-7} 

& \multirow{5}{*}{Gamma-shift} 

& Gamma-padding & 60.2 & 40.6 
& \good{-19.60} & \good{-32.56\%} \\ 

& & \textbf{Gamma-shift} & \textbf{99.6} & \textbf{95.9} 
& \textbf{\good{-3.70}} & \textbf{\good{-3.71\%}} \\ 

& & MAB-malware & 65.1 & 64.0 
& \good{-1.10} & \good{-1.69\%} \\

& & MiniMal & 90.8 & 66.6 
& \good{-24.20} & \good{-26.65\%} \\

& & AIMED & 29.2 & 42 
& \bad{12.80} & \bad{43.84\%} \\ 

\cmidrule(lr){2-7} 

& \multirow{5}{*}{MAB-malware} 

& Gamma-padding & 60.2 & 40.5 
& \good{-19.70} & \good{-32.72\%} \\ 

& & Gamma-shift & 99.6 & 99.7 
& \bad{0.10} & \bad{0.10\%} \\ 

& & \textbf{MAB-malware} & \textbf{65.1} & \textbf{59.7} 
& \textbf{\good{-5.40}} & \textbf{\good{-8.29\%}} \\ 

& & MiniMal & 90.8 & 56.7 
& \good{-34.10} & \good{-37.56\%} \\

& & AIMED & 29.2 & 36.6 
& \bad{7.40} & \bad{25.34\%} \\ 

\cmidrule(lr){2-7} 

& \multirow{5}{*}{MiniMal} 

& Gamma-padding & 60.2 & 51.96 
& \good{-8.24} & \good{-13.69\%} \\ 

& & Gamma-shift & 99.6 & 99.55 
& \good{-0.05} & \good{-0.05\%} \\ 

& & MAB-malware & 65.1 & 60.33 
& \good{-4.77} & \good{-7.33\%} \\ 

& & \textbf{MiniMal} & \textbf{90.8} & \textbf{51.3} 
& \textbf{\good{-39.50}} & \textbf{\good{-43.50\%}} \\ 

& & AIMED & 29.2 & 9.5 
& \good{-19.70} & \good{-67.47\%} \\ 

\cmidrule(lr){2-7} 

& \multirow{5}{*}{AIMED} 

& Gamma-padding & 60.2 & 59.6 
& \good{-0.60} & \good{-1.00\%} \\ 

& & Gamma-shift & 99.6 & 99.7 
& \bad{0.10} & \bad{0.10\%} \\ 

& & MAB-malware & 65.1 & 68.2 
& \bad{3.10} & \bad{4.76\%} \\ 

& & MiniMal & 90.8 & 64.9 
& \good{-25.90} & \good{-28.52\%} \\

& & \textbf{AIMED} & \textbf{29.2} & \textbf{12.7} 
& \textbf{\good{-16.50}} & \textbf{\good{-56.51\%}} \\ 

\cmidrule(lr){1-7}

\multirow{25}{*}{\textbf{Commercial}} 
& \multirow{5}{*}{Gamma-padding} 

& \textbf{Gamma-padding} & \textbf{93} & \textbf{7.5} 
& \textbf{\good{-85.50}} & \textbf{\good{-91.94\%}} \\ 

& & Gamma-shift & 94 & 28.9 
& \good{-65.10} & \good{-69.26\%} \\ 

& & MAB-malware & 99.1 & 81.8 
& \good{-17.30} & \good{-17.46\%} \\ 

& & MiniMal & 98.5 & 81.8 
& \good{-16.70} & \good{-16.95\%} \\ 

& & AIMED & 33.2 & 35.9 
& \bad{2.70} & \bad{8.13\%} \\ 

\cmidrule(lr){2-7} 

& \multirow{5}{*}{Gamma-shift} 

& Gamma-padding & 93 & 21.1 
& \good{-71.90} & \good{-77.31\%} \\ 

& & \textbf{Gamma-shift} & \textbf{94} & \textbf{5.9} 
& \textbf{\good{-88.10}} & \textbf{\good{-93.72\%}} \\ 

& & MAB-malware & 99.1 & 83.8 
& \good{-15.30} & \good{-15.44\%} \\ 

& & MiniMal & 98.5 & 83.8 
& \good{-14.70} & \good{-14.92\%} \\ 

& & AIMED & 33.2 & 36.5 
& \bad{3.30} & \bad{9.94\%} \\ 

\cmidrule(lr){2-7} 

& \multirow{5}{*}{MAB-malware} 

& Gamma-padding & 93 & 81.6 
& \good{-11.40} & \good{-12.26\%} \\ 

& & Gamma-shift & 94 & 90 
& \good{-4.00} & \good{-4.26\%} \\ 

& & \textbf{MAB-malware} & \textbf{99.1} & \textbf{99.6} 
& \textbf{\bad{0.50}} & \textbf{\bad{0.50\%}} \\ 

& & MiniMal & 98.5 & 72.0 
& \good{-26.50} & \good{-26.90\%} \\ 

& & AIMED & 33.2 & 38.5 
& \bad{5.30} & \bad{15.96\%} \\ 

\cmidrule(lr){2-7} 

& \multirow{5}{*}{MiniMal} 

& Gamma-padding & 93 & 91.83 
& \good{-1.17} & \good{-1.26\%} \\ 

& & Gamma-shift & 94 & 98.29 
& \bad{4.29} & \bad{4.56\%} \\ 

& & MAB-malware & 99.1 & 100 
& \bad{0.90} & \bad{0.91\%} \\ 

& & \textbf{MiniMal} & \textbf{98.5} & \textbf{77.5} 
& \textbf{\good{-21.00}} & \textbf{\good{-21.32\%}} \\ 

& & AIMED & 33.2 & 16.5 
& \good{-16.70} & \good{-50.30\%} \\ 

\cmidrule(lr){2-7} 

& \multirow{5}{*}{AIMED} 

& Gamma-padding & 93 & 97.4 
& \bad{4.40} & \bad{4.73\%} \\ 

& & Gamma-shift & 94 & 96.8 
& \bad{2.80} & \bad{2.98\%} \\ 

& & MAB-malware & 99.1 & 99.4 
& \bad{0.30} & \bad{0.30\%} \\ 

& & MiniMal & 98.5 & 86.4 
& \good{-12.10} & \good{-12.28\%} \\ 

& & \textbf{AIMED} & \textbf{33.2} & \textbf{24.3} 
& \textbf{\good{-8.90}} & \textbf{\good{-26.81\%}} \\ 

\bottomrule 
\end{tabular}
}
\end{table*}